\documentclass[%
 reprint,
 amsmath,amssymb,
 aps, prx,
]{revtex4-2}

\usepackage{amsmath}
\usepackage[table,xcdraw]{xcolor}
\usepackage{graphicx}% Include figure files
\usepackage{dcolumn}% Align table columns on decimal point
\usepackage{bm}% bold math
\usepackage{float}
\usepackage{dsfont}

\usepackage{algpseudocode}
\usepackage{algorithm}
\usepackage{physics}
\usepackage{color}
\usepackage{placeins}
\usepackage[colorlinks=true, linkcolor=blue, citecolor=blue, urlcolor=blue]{hyperref}

\usepackage[capitalise]{cleveref}
\usepackage[all]{hypcap}
\usepackage{booktabs} 
\usepackage{amsmath} 						% Formelumgebungen
\usepackage{amssymb} 						% Mathe-Symbole
\usepackage{mathtools}			% Erweiterung von amsmath
\usepackage{mathrsfs}                       % script mathlanguage
\usepackage{amsthm}
\usepackage{amssymb}
\usepackage{hyperref}

\usepackage{quantikz}     % Quantum circuit diagrams
\usepackage{amsmath}      % Math
\usepackage{amssymb}
\usepackage{graphicx}
\usepackage{booktabs}  % Add this in your preamble

\newcommand{\sy}{\sigma_{\text{y}}}
\newcommand{\sz}{\sigma_{\text{z}}}
\newcommand{\splus}{\sigma_{+}}
\newcommand{\sm}{\sigma_{-}}

\newcommand{\ii}{\text{i}}
\renewcommand{\Tr}{\text{Tr}}

\begin{document}

\preprint{APS/123-QED}

%\title{Quantum Reservoir Computing maps Data onto the Krylov Space}

\title{Memory-Nonlinearity Trade-off across Quantum Reservoir Computing Frameworks}

%\title{Quantum Reservoir Computing maps Data onto the Krylov Space}
 
\author{Saud \v{C}indrak$^{1}$}
\email{saud.cindrak@tu-ilmenau.de}
\author{Lara Giebeler$^{2}$}
\author{Niclas Götting$^{3}$}
\author{Christopher Gies$^{3}$}
\author{Kathy Lüdge$^{1}$}

\affiliation{$^{1}$Institute of Physics, Technische Universit\"at Ilmenau, Ilmenau, Germany}

\affiliation{$^{2}$Institut für Physik und Astronomie, Technische Universität Berlin, 
Hardenbergstr.\ 36, D-10623 Berlin, Germany}

\affiliation{$^{3}$Carl von Ossietzky Universität Oldenburg, Fakultät V, Institut für Physik, 
26129 Oldenburg, Germany}

\date{\today}

\begin{abstract}
 
Quantum reservoir computing (QRC) harnesses driven quantum dynamics for time-series processing, yet the mechanisms behind the differing performance levels across its many implementations remain unclear. 
We show that apparently unrelated approaches—including memory restriction, weak measurements, operation near the edge of quantum chaos, and dissipative dynamics—are in fact governed by the same underlying principle, namely a tunable balance between memory retention and nonlinear response.
Using the information processing capacity, a dynamical measure from nonlinear systems theory, we place these behaviors in a unified framework and identify the regimes in which quantum reservoirs surpass the standard protocol. 
Our results reveal a fundamental connection between memory and nonlinear response. 
This provides a general design principle for enhanced information processing and enables systematic analysis and optimization inspired by classical dynamical quantifiers.

\end{abstract}

\maketitle

\section{Introduction}
Recent years have seen significant interest in quantum computing and machine learning from both industry and academia. While quantum computing promises to solve tasks that are inaccessible to classical computers \cite{NIE06a}, machine learning is actively transforming our society. In this context, the field of quantum machine learning raises the question of whether advantages can be achieved by running machine-learning models on quantum hardware \cite{CER22}. The promise behind such an approach is that quantum algorithms do not need to be manually designed; instead, machine-learning frameworks can adjust the necessary operations to achieve quantum advantage.

Many proposals exist for implementing quantum machine learning, ranging from variational quantum circuits \cite{ARU19, SCH15a, BIA17, CER22} and quantum kernel methods \cite{HAV19, SCH21a} to quantum reservoir computing (QRC) \cite{FUJ17}, the latter being the focus of this paper. QRC, first introduced by Fujii and Nakajima \cite{FUJ17}, takes after classical reservoir computing \cite{JAE01}, where physical systems are exploited for computation.
In its simplest form, a time-series input is encoded into a quantum state; the system then evolves naturally under quantum dynamics, and the expectation values of a set of observables are measured and form the output. Training is performed via linear regression on these outputs. This approach has gained significant attention in recent years, as demonstrated by various research directions exploring different aspects of the framework \cite{SUZ22, NAK19, GAR23, MUJ21a, ABB24a, FUJ21, NAK20, PFE23, KAW26, KOB24a}. 
In particular, substantial progress has been made in the interplay of QRC and quantum mechanics, including investigations of entanglement \cite{GOE23}, operator growth \cite{VET25}, expressivity \cite{XIO25} and Krylov complexity \cite{DOM24, CIN24, CIN25a, CIN25}. These analyses quantify different quantum-mechanical features, reveal their influence on computational performance, and shed light on the interplay between data processing and quantum dynamics.

In principle, the design of quantum reservoirs can be separated into input encoding, time evolution, and measurement, each leading to different considerations and challenges. One such challenge is the collapse of the state during measurement. Because QRC typically processes time-series inputs, measurement destroys information about previous inputs, resulting in information loss and the need to re-encode the full time series up to that point, a procedure referred to as the fully restarting protocol (FRP) \cite{FUJ17}. One way to mitigate this limitation is to use a weak measurement protocol  (WMP) \cite{MUJ23}, utilizing the reduced amount of measurement back-action on the system to avoid restarts. Recently, it was also shown that in some cases this even improves task performance \cite{FRA24}. Another approach exploits the inherent fading-memory property of quantum reservoirs discussed in \cite{CHE24b}, proposing to restrict computation only to the relevant past inputs \cite{MUJ23, CIN24}. This method, often referred to as the memory-restricted protocol (MRP), can also exhibit improved performance compared to the standard FRP protocol. A recurrence-free QRC can also be implemented by re-encoding the measured output, resulting in a feedback-driven QRC scheme \cite{PFE22, AHM24, KOB24}.

% BEFORE this, you are in two-column mode
\begin{figure*}[t]
    %================ LEFT: (a) Protocol sketches ================
    \begin{minipage}[t]{0.48\textwidth}
        \raggedright  % left-align inside this block

        \textbf{(a) QRC Frameworks}\\[2mm]

        %---- Restarting Protocol ----
        \vspace{6mm}
        \textbf{Full restarting protocol (FRP)}\\[1mm]
        {\fontsize{8}{8}\selectfont
        \begin{quantikz}[column sep=0.2cm]
            \lstick{$\rho_0$} & \gate{U_E(u_1)} & \gate{U_R} &
                \gate{U(u_2)} & \gate{U_R} & \push{\dots} &
                \gate{U(u_n)} & \gate{U_R} & \meter{}
        \end{quantikz}
        }

        \vspace{6mm}

        %---- Restricted Protocol ----
        \textbf{Memory-restricted protocol (MRP)}\\[1mm]
        {\fontsize{8}{8}\selectfont
        \begin{quantikz}[column sep=0.2cm]
            \lstick{$\rho_0$} & \gate{U_E(u_{n-r})} & \gate{U_R} &
                \push{\dots}\push{\dots} &
                \gate{U(u_n)} & \gate{U_R} & \meter{}
        \end{quantikz}
        }

        \vspace{6mm}

        %---- Weak Measurements ----
        \textbf{Weak measurements protocol (WMP)}\\[1mm]
        {\fontsize{8}{8}\selectfont
        \begin{quantikz}[column sep=0.2cm]
            \lstick{$\rho_0$} & \gate{U_E(u_1)} & \gate{U_R} &
                \meter{\vartheta} &
                \push{\dots}\push{\dots}  &
                \gate{U(u_n)} & \gate{U_R} & \meter{\vartheta}
        \end{quantikz}
        }

        \vspace{6mm}

        %---- Dissipation ----
        \textbf{Dissipative protocol (DSP)}\\[1mm]
        {\fontsize{8}{8}\selectfont
        \begin{quantikz}[column sep=0.2cm]
            \lstick{$\rho_0$} & \gate{e^{\mathcal{L}(u_1, \gamma)}} & \gate{e^{\mathcal{L}(u_2, \gamma)}} &
                \push{\dots} \push{\dots} &
                 \gate{e^{\mathcal{L}(u_n, \gamma)}} &
                \meter{}
        \end{quantikz}
        }

    \end{minipage}
    \hfill
    %================ RIGHT: (b) Results plot =====================
    \begin{minipage}[t]{0.48\textwidth}
        \raggedright

        \textbf{    (b) Main Result}\\[2mm]

        \includegraphics[scale = 1]{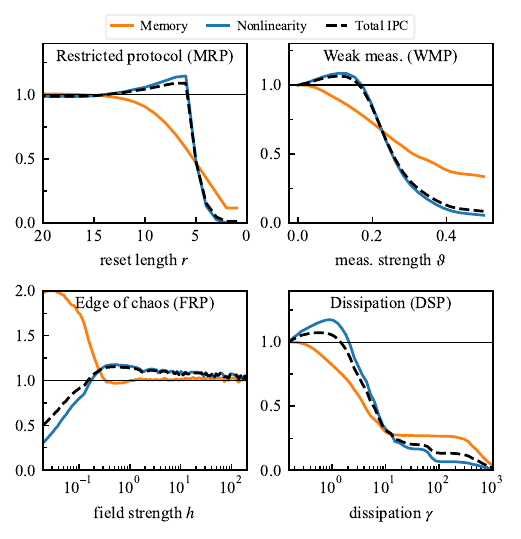}

        % Optionally add a short description here:
        % {\fontsize{8}{8}\selectfont
        % Example task performance for the different protocols.
        % }
    \end{minipage}

    \caption{(a) Schematic overview of the quantum reservoir computing protocols: restarting, restricted input, weak measurement, and dissipative variants. 
    (b) Example performance results obtained with these protocols on a representative task, where the restarting protocol is used to compute the edge of chaos. 
    Definitions of memory, nonlinearity, and the total IPC are given in \cref{subseq:IPC}. 
    We normalize the values in (a) and (b) to the FRP protocol, in (c) to the ergodic case, which would correspond to a random topology in classical reservoir computing, and in (d) to the maximum of linearity.}
\label{fig:protocols_results}
\end{figure*}

A further aspect concerns the feasibility of running quantum reservoirs on current hardware, which exhibit dissipation through interaction with the environment \cite{CHE20b, SAN24, DOM23, FRY23, GOE25}. Interestingly, these works suggest that dissipation can serve as a computational resource. Lastly, the nature of the quantum system itself can serve as an indicator of optimal task performance, as studies suggest that ideal performance occurs at the edge of quantum chaos \cite{MAR21, KOB26}.

Although these performance enhancements have been analyzed through various quantum-mechanical concepts, there is currently no unifying framework that explains how input encoding, time evolution, reservoir structure, and measurement observables influence task performance.
To move towards such a unifying perspective, it is natural to complement quantum-mechanical analyses with tools from nonlinear dynamical systems. One promising direction is the characterization of QRC using dynamical quantifiers, such as the information processing capacity, which measures how well a system can retain and nonlinearly transform past input perturbations \cite{DAM12}. This framework has provided deep insight into the temporal processing capabilities of a wide range of classical models and reservoirs \cite{BAU24, HUE22a, INU17, RAT25, HUA26, AUS25}. In the quantum setting, \cite{MAR20} first analyzed how multiplexing, system size, and measurement directions affect this quantity, while in \cite{CIN24} we studied the information processing capacity of the restricted protocol.  \\

Building on the observed performance improvements across various QRC frameworks—including restricted and restarting protocols \cite{FUJ17, MUJ23, CIN24}, weak measurements \cite{MUJ23, FRA24}, phase transitions and quantum chaos \cite{MAR21, KOB26}, and dissipation-based approaches \cite{CHE20b, SAN24, DOM23, FRY23, GOE25}—we hypothesize that these gains are fundamentally rooted in the same underlying principle: a yet unexplored memory–nonlinearity trade-off in QRC. Furthermore, we demonstrate that each of these approaches exhibits regions of enhanced information processing compared to the standard QRC model, paralleling classical results \cite{INU17}. We quantify this trade-off by computing the information processing capacity and compare it across three benchmark tasks: Lorenz future prediction, Lorenz cross prediction, and the Mackey--Glass task \cite{LOR63, WER19}. 
Our results support this hypothesis. 

In \cref{subseq:main_result}, we provide a brief overview of the main results of this work, revealing a striking similarity across all protocols when comparing the memory and nonlinearity of the QRC protocols. This trade-off can be understood by considering the Hilbert space in which the operators evolve and by using arguments from the memory-restriction protocol. 
\cite{CIN25, CIN25a} have shown that the degrees of freedom of the quantum evolution are determined by the Hamiltonian and the measured observables, and are therefore independent of the input encoding. Restricting the number of inputs through one of the discussed QRC protocols maps fewer inputs onto these fixed degrees of freedom. This enables more complex mixtures between inputs, as also observed in \cite{CIN24}, which ultimately leads to the observed memory--nonlinearity trade-off.

This work is organized as follows. In \cref{sec:qrc_theory} we provide an overview of the standard quantum reservoir computing (QRC) framework \cite{FUJ17}, referred to as the restarting protocol, and introduce both the information-processing capacity (IPC) formalism \cite{DAM12} and the time-series prediction tasks considered in this study \cite{LOR63}.
We then discuss the task performance and the memory--nonlinearity trade-off, as quantified by the IPCs, for the memory-restriction protocol (MRP, \cref{sec:restricted}), the weak-measurement approach (\cref{sec:weak}), the behavior at the phase transition (\cref{sec:chaos}), and dissipative dynamics governed by Lindblad evolution (\cref{sec:lindblad}).
Our results show that every QRC framework exhibits the inherent memory–nonlinearity trade-off discussed above, demonstrating this as a guiding principle and a missing link in the current QRC literature.

\subsection{Main Result}\label{subseq:main_result}

To give the reader a unified perspective on the main results, \cref{fig:protocols_results}(a) illustrates the QRC frameworks analyzed in this work: the fully restarting protocol (FRP), the memory-restricted protocol (MRP), the weak-measurement protocol  (WMP), and a dissipative protocol (DSP). We note that, for the analysis of phase transitions, we employ the restarting protocol and sweep a Hamiltonian parameter accordingly.
For each of these QRC frameworks, we compute the information-processing capacities (IPCs) up to sixth order and plot the first-order contribution $\mathrm{IPC}_1$ (orange), also referred to as the memory capacity, the nonlinear contributions $\mathrm{IPC}_{\geq 2}$ (blue), and the sum of both memory capacity and nonlinear response (black, dashed) in \cref{fig:protocols_results}(b). 
We normalize the subplots in \cref{fig:protocols_results}(b) as follows: For MRP and WMP, the results are normalized with respect to the FRP. For the edge of chaos, we normalize with respect to a Hamiltonian deep in the ergodic region, and for the DSP we normalize with respect to the point of maximal memory.

We first analyze the MRP, which re-encodes only the most recent $r$ inputs at each timestep (see \cref{fig:protocols_results}(a), second row). In the limit where the full input time series is re-encoded, the MRP naturally reduces to the FRP, shown in \cref{fig:protocols_results}(a), first row. Accordingly, in \cref{fig:protocols_results}(b) (upper left), we present the memory retention, nonlinear response, and total IPC for the restarting protocol as functions of decreasing reset length $r$. The results are normalized to the FRP values, and for $r = 20$ we observe that MRP and FRP yield identical memory, nonlinearity, and total IPC.
As expected, decreasing the number of encoded inputs results in a reduction of the system's memory retention (orange). Conversely, we observe an increase in the nonlinear response (blue), demonstrating the hypothesized memory--nonlinearity trade-off in QRC. When considering the sum of memory retention and nonlinear response (black, dashed), we find that the MRP protocol exceeds the FRP protocol, as indicated by normalized values greater than one. These results reveal both an inherent memory--nonlinearity trade-off in the MRP protocol and a regime in which the processing capabilities surpass this trade-off. This highlights that memory retention can act as a resource to enhance the information-processing capacities of quantum reservoirs.
We provide a detailed analysis of this behavior in \cref{sec:restricted}.

Next we analyze the effect of performing mid-process measurements in a weak measurement protocol as shown in \cref{fig:protocols_results}(a) in the third row, by characterizing their impact on the expressivity and the task-specific performance of the WMP. At one extreme, a strong measurement with a high measurement strength $\vartheta$ collapses the state and extracts the required information. At the other extreme a weak measurement extracts only limited information but leaves the state mostly undisturbed. Our results show that varying the measurement strength is an additional way to tune the memory-nonlinearity trade-off in QRC. As shown in \cref{fig:protocols_results}(b) (upper right),  stronger measurements introduce more back-action into the system, therefore shifting the trade-off towards nonlinear processing capacities (blue). For very weak measurements we recover the limit of the FRP and maximize the reservoir's memory retention (yellow), since no additional back-action is introduced by the measurements. We again find a sweet spot, where the total processing capacity (black, dashed) is maximized, surpassing the FRP limit, by optimizing the memory-nonlinearity trade-off. In \cref{sec:weak}, we analyze this in more detail and show that by tuning this trade-off we can optimize task-specific performance.%, which in some cases requires more memory and in others more nonlinear processing capacity.

To examine how the memory--nonlinearity trade-off depends on the dynamical phase of the reservoir, we compute the memory capacity, nonlinear response, and total IPC as functions of the Hamiltonian's transverse field strength $h$ (see \cref{fig:protocols_results}(b), lower left). For small $h<0.1$ the system exhibits strong localization, while for $h>2$ we can assume the system to be chaotic, often also addressed as ergodic in the quantum dynamics literature \cite{MAR21, KOB26}. The values of the IPCs in \cref{fig:protocols_results}(b) are normalized to the deep ergodic case at $h = 10^{2}$.
We motivate this normalization through classical reservoir computing, where the internal weights are sampled from a random distribution. Such internal weights exhibit eigenvalue statistics similar to those of a Hamiltonian in the deeply ergodic phase~\cite{MAR21}. Thus, comparing a quantum-mechanical model such as the Ising Hamiltonian to its ergodic regime provides a natural reference point. With this normalization in place, we now examine how the reservoir’s computational features evolve across dynamical phases. For small field strengths the system lies in a localized phase, where memory retention is up to a factor of two larger than in the ergodic regime reached at large~$h$. The region where the nonlinear terms exceed their ergodic value appears near $h \approx 0.5$, marking the phase transition. In this critical regime, the nonlinear contributions exhibit a clear maximum, while memory retention decreases below the ergodic limit. Recent studies suggest that the region of maximal quantum chaos occurs at the phase transition, which interestingly overlaps with the largest nonlinear response of the system \cite{SWI25}.
The resulting enhancement of the nonlinear response therefore leads to a larger total information-processing capacity, as indicated by the black dashed curve rising above one. These observations highlight the presence of a memory--nonlinearity trade-off across the phase transition and underscore the importance of systematically and physically informed reservoir design. Further analysis of the individual IPC components and their impact on task performance is provided in \cref{sec:chaos}.

Finally, in \cref{sec:lindblad} we examine the relation between memory and nonlinearity in quantum reservoirs governed by dissipative Lindblad dynamics, as shown in the bottom row of \cref{fig:protocols_results}(a).
%We normalize all results at the parameter value where $\mathrm{IPC}_1(\gamma)$ attains its maximum.
In contrast to the unitary input mechanisms discussed above, fading memory appears naturally through qubit decay, eliminating the need for a state-reset input encoding; instead, an amplitude-controlled coherent drive is employed.
When dissipation is very small, the system dynamics fails to display fading memory, which causes inputs to accumulate and reduces task performance (see \cref{sec:lindblad} for more details). 
As the decay rate $\gamma$ increases, fading memory improves while nonlinear processing remains weak at first as illustrated in the bottom right panel of \cref{fig:protocols_results}(b).
At intermediate dissipation, memory retention and nonlinear response become well balanced, leading to the highest IPC.
Further improvement in nonlinear processing capacity for even faster dissipation rates, however, leads to reduced linear memory.
%Interestingly, the intermediate regime also gives the best performance on the Lorenz tasks, showing that controlled dissipation offers an effective method for tuning both memory and nonlinear response.
For larger values of $\gamma$, memory is lost rapidly and computation fails, with lower order IPC terms diminishing first.

When comparing the normalized memory, nonlinearity, and their sum for the restricted protocol, weak measurement, edge-of-chaos dynamics, and dissipative regimes (see \cref{fig:protocols_results}(b)), a clear similarity emerges, hinting at a universal behavior when these implementations are analyzed through the lens of the memory--nonlinearity trade-off. For each of these QRC frameworks, a reduction in memory initially leads to an increase in nonlinearity, illustrating the inherent memory--nonlinearity trade-off. Furthermore, each framework exhibits a regime in which the total information processing capacity surpasses the limitations imposed by this intrinsic trade-off. For the restricted protocol, this regime occurs at intermediate reset lengths ($r = 6$); for weak measurements at $\vartheta \approx 0.11$; near the phase transition at $h \approx 0.5$; and for dissipative dynamics at $\gamma \approx 10^{0}$.
 
%In summary, this work provides a unified analysis of diverse quantum reservoir computing architectures through the lens of the memory–nonlinearity trade-off and shows information processing beyond this trade-off. By evaluating weak measurements, memory restriction, phase transitions and dissipation within a common framework, we demonstrate that all these mechanisms exhibit the same fundamental structure: enhanced nonlinear processing accompanied by controlled memory depth. Through systematic computation of the IPC and benchmarking across Lorenz future prediction, Lorenz cross prediction, and Mackey–Glass forecasting, we show that performance improvements observed throughout the QRC literature can be consistently traced back to this universal trade-off. Our results thus establish the memory–nonlinearity balance as a central design principle of quantum reservoir computing and offer a coherent theoretical explanation for performance gains across otherwise distinct QRC frameworks.

\section{Quantum Reservoir Computing} \label{sec:qrc_theory}

In general, the unperturbed time evolution of a quantum reservoir using the Fujii Nakajima map \cite{FUJ17} can be separated into input encoding, time evolution governed by a quantum system described by a Hamiltonian \(H\), and a set of measured observables, in our case \(\sigma_z^{(k)}\). The sequence \(\{(u_k, f_k)\}_{k \in \mathbb{N}}\) represents the input–target pairs, where the input \(u_k\) is encoded into the encoding qubits. We consider a one-dimensional time series whose inputs are encoded into the first qubit as
\begin{align}
\rho_{1}(u_k) = \sqrt{\frac{1 + u_k}{2}} \ket{0} + \sqrt{\frac{1 - u_k}{2}} \ket{1}
\end{align}
and the encoding is adapted to support inputs \(u_k \in [-1,1]\). The state is updated according to
\begin{align}
\rho(s_k) = U_R \bigl( \rho_{1}(u_k) \otimes \mathrm{Tr}_{1}[\rho(s_{k-1})] \bigr) U_R^\dagger.
\end{align}
\(\mathrm{Tr}_{1}\) denoting the partial trace over the first qubit applied to the previous state \(\rho(s_{k-1})\). The operator \(U_R = \exp(-i H T)\) represents the unitary time evolution. The parameter \(T\) is the time for which the system evolves before the next input is encoded and is typically referred to as the clock cycle.

Throughout this study, we consider quantum reservoirs whose dynamics are governed by the fully connected transverse-field Ising model (TFIM)
\begin{align}
H = \frac{1}{2} \sum_{i=1}^{N_S} h \sigma_{\mathrm{z}}^{(i)} + \sum_{i<j}^{N_S} J_{ij} \sigma_{\mathrm{x}}^{(i)} \sigma_{\mathrm{x}}^{(j)}.
\label{eq:ising_chaos}
\end{align}
\(\sigma_{\mathrm{z}}^{(i)}\) and \(\sigma_{\mathrm{x}}^{(i)}\) represent the Pauli z and Pauli x matrices operators on the \(i\)th qubit of an \(N_S\)-qubit system. We sample \(J_{ij} \in \mathcal{U}([0,1])\) and set \(h = 1\), which represents operation at the phase transition and has been shown to exhibit optimal performance \cite{MAR21, KOB26}. We further normalize the Hamiltonian such that the spectral radius satisfies \(\rho(H) = 1\).

The readout \({x}(u_k)\) is a vector obtained by computing the expectation values of the Pauli z observables \(\sigma_{\mathrm{z}}^{(i)}\) at each site
\begin{align}
    {x}(u_k) = \bigl[ \mathrm{Tr}[\rho(s_k)\sigma_{\mathrm{z}}^{(1)}], \ldots, \mathrm{Tr}[\rho(s_k)\sigma_{\mathrm{z}}^{(N_S)}] \bigr],
\end{align}
resulting in a readout dimension of \(N_S\). 
To further increase the readout dimension we measure the expectation values \(N_V\) times at \(\tau_k = k T / N_V\). This produces a readout dimension of \(N_R = N_S N_V\). In this work, we consider $N_S = 4$ and $N_V = 30$, and a clock cycle of $T = 50$, unless stated otherwise. The output vector can, therefore, be written as
\begin{align}
    {x}(u_k) = \bigl[ x_1(u_k), x_2(u_k), \ldots, x_{N_R}(u_k) \bigr].
\end{align}
The probabilistic nature of the measurement process is simulated by sampling from a Gaussian distribution,
\begin{align}
\langle \sigma_z^{(i)} \rangle \sim \mathcal{N}\left( \mathrm{Tr}[\sigma_z^{(i)} \rho], \frac{1}{N_{\mathrm{meas}}} \right),
\label{eq:shot_noise}
\end{align}
where the shot noise standard deviation is 
\(\sigma_{\mathcal{N}} = 1 / \sqrt{N_{\mathrm{meas}}} = 10^{-5}\), 
which is used throughout this study.
The output \(\hat{f}\) of the system is constructed by a linear estimator
\begin{align}
\hat{f}(u_k) = \sum_{i=1}^{N} w_i x_i(u_k),
\end{align}
where \(w_i\) are the output weights trained to minimize the quadratic loss
\begin{align}
\mathrm{MSE}[\hat{f}, f] = \frac{1}{K} \sum_{k=1}^K \bigl( \hat{f}(u_k) - f(u_k) \bigr)^2
\end{align}
with \(f(u_k)\) representing the target. To compute the weights, we consider a time series \((u_k)_{k=1}^{N_{\mathrm{Tr}}}\) with targets \(f(u_k)\) and collect the state matrix
\begin{align}
\mathbf{X} = \bigl[ x(u_1), x(u_2), \ldots, x(u_{N_{\mathrm{Tr}}}) \bigr]^{\mathrm{T}}.
\end{align}
The weights are then given by
\begin{align}
    w = (\mathbf{X}^{\mathrm{T}}\mathbf{X})^{-1} \mathbf{X}^{\mathrm{T}} {f},
\end{align}
where \({f} = [f(u_1), f(u_2), \ldots, f(u_{N_{\mathrm{Tr}}})]^{\mathrm{T}}\). Throughout this work, we use training and test sets of sizes $N_{\mathrm{tr}} = 50\,000$ and $N_{\mathrm{te}} = 5\,000$, respectively.
The normalized mean squared error analyzed throughout this study is
\begin{align}
\mathrm{NRMSE}[\hat{f}, f]
=
\sqrt{
\frac{\mathrm{MSE}[\hat{f}, f]}{\mathrm{Var}(f)}
}.
\label{eq:nrmse}
\end{align}
We further note that the shot noise inherent in the quantum dynamics (\cref{eq:shot_noise}) naturally serves as a regularization parameter, in line with the concept of regularization by noise \cite{BIS95}, and was used in QRC in our previous works \cite{CIN24, CIN25, CIN25a}.

\subsection{Information Processing Capacity}\label{subseq:IPC}

The information processing capacity (IPC) quantifies how well a system can retain and nonlinearly transform past inputs \cite{DAM12}. It has been analyzed for different input distributions \cite{KUB21} and studied in relation to the eigenvalue spectrum of the reservoir dynamics \cite{KOE21, KOE22}. In QRC, it has been shown that the total IPC correlates almost perfectly with the degrees of freedom captured by operator growth as quantified by Krylov observability \cite{CIN25, CIN25a}. 

In this work, the input is assumed to be uniformly distributed on the interval $u \in \mathcal{U}([-1,1])$. For this distribution, the appropriate orthogonal basis is given by the Legendre polynomials, defined by
\begin{align}
    l_0(x) &= 1,  \qquad l_1(x) = x,  \qquad 
    l_2(x) = \tfrac{1}{2}(3x^2 - 1), \nonumber\\
    l_n(x) &= \frac{2n - 1}{n} \, x\, l_{n-1}(x) - \frac{n - 1}{n}\, l_{n-2}(x), \quad n \ge 2.
\end{align}
Using these polynomials, IPC target functions are constructed by evaluating the Legendre polynomials on time-shifted inputs. For a time delay $d$ and an input sequence $(u_k)_{k=1}^{N_u}$, the target corresponding to the $k$-th Legendre polynomial is
\begin{align}
    l^{(k)}_d = \big[\, l_k(u_{1-d}),\; l_k(u_{2-d}),\; \ldots,\; l_k(u_{N_u-d}) \,\big]^T.
\end{align}
Higher-order targets are formed by point-wise products of multiple Legendre-polynomial targets,
\begin{align}
    f^{(k_1, k_2, \ldots, k_n)}_{d_1, d_2, \ldots, d_n}
    = \bigodot_{i=1}^{n} l^{(k_i)}_{d_i},
\end{align}
where $\bigodot$ represents element-wise multiplication.

\medskip
\noindent\textbf{First-order capacity.}
The first-order IPC ($\mathrm{IPC}_1$) is defined through targets associated with the first Legendre polynomial. Since $l_1(x)=x$, these targets reduce to
\begin{align}
    f_d^{(1)} = \big[\, u_{1-d},\; u_{2-d},\; \ldots,\; u_{N_u-d} \,\big]^T
\end{align}
and evaluate how accurately the reservoir reproduces delayed input values.  
Consequently, $\mathrm{IPC}_1$ is often identified with the memory capacity \cite{GOL20, KOE20a, BAU22c}.
For each delay $d$, the reservoir produces a prediction $\hat{f}_d^{(1)}$ after training, and the capacity is defined as
\begin{align}
    C\!\left(f_d^{(1)}, \hat{f}_d^{(1)}\right) 
        = \left(
            \frac{\operatorname{cov}\!\left(f_d^{(1)},\, \hat{f}_d^{(1)}\right)}
                 {\sigma\!\left(f_d^{(1)}\right)\,\sigma\!\left(\hat{f}_d^{(1)}\right)}
          \right)^2,
\end{align}
where $\operatorname{cov}$ and $\sigma$ denote covariance and standard deviation.

\medskip
\noindent\textbf{Second-order capacity.}
The second-order IPC (IPC$_2$) consists of two types of target functions.  
Pure second-order targets are given by
\begin{align}
    f_d^{(2)} = \big[\, l_2(u_{1-d}),\;  \ldots,\; l_2(u_{N_u-d}) \,\big]^T = l_d^{(2)},
\end{align}
and mixed targets are formed by products of two first-order terms,
\begin{align}
    f_{d_1,d_2}^{(1,1)} 
    = l_{d_1}^{(1)}\odot l_{d_2}^{(1)} %big[\, l_1(u_{1-d_1})\,l_1(u_{1-d_2}),\; \ldots,\; l_1(u_{N_u-d_1})\,l_1(u_{N_u-d_2}) \,\big]^T,
\end{align}
with the constraint $d_1>d_2$ to avoid redundant combinations.  
The total second-order capacity is
\begin{align}
    \mathrm{IPC}_{2,1} &= \sum_{d} C\!\left(f_d^{(2)},\, \hat{f}_d^{(2)}\right), \\
    \mathrm{IPC}_{2,2} &= \sum_{\substack{d_1,d_2\\ d_1>d_2}}
        C\!\left(f_{d_1,d_2}^{(1,1)},\, \hat{f}_{d_1,d_2}^{(1,1)}\right), \\
    \mathrm{IPC}_2 &= \mathrm{IPC}_{2,1} + \mathrm{IPC}_{2,2}.
\end{align}

\medskip
\noindent\textbf{Higher-order capacities.}
For order $D$, all Legendre-polynomial combinations with total degree $\sum_j d_i = D$ are used as targets. For example, third-order ($D=3$) IPC includes
\begin{align}
    f^{(3)}_{d} &= l^{(3)}_{d}  \nonumber\\
    f^{(1,2)}_{d_1,d_2} &= l^{(1)}_{d_1}\odot l^{(2)}_{d_2}, \nonumber\\
    f^{(1,1,1)}_{d_1,d_2,d_3} &=  l^{(1)}_{d_1}\odot l^{(1)}_{d_2}\odot l^{(1)}_{d_3},
\end{align}
summed over all valid index combinations with ordering constraints to avoid duplicates. A detailed discussion of how these combinations interact across orders and delays is provided in \cite{JAE01, CIN24, HUE22a, KOE20a}.

\medskip
\noindent\textbf{Total IPC.}
The total information processing capacity is obtained by summing over all orders,
\begin{align}
    \mathrm{IPC} = \sum_i \mathrm{IPC}_i,
\end{align}
and is upper-bounded by the readout dimension $N_R$ \cite{DAM12}. We note that, in the computation of the IPC, we apply a cutoff below which all capacity values are set to zero. This cutoff is determined through the shuffling method provided in \cite{KUB21}.

\subsection{Benchmarks}
Using an appropriate benchmark to evaluate performance is as important as the algorithm used itself and is therefore still a great challenge in (quantum) machine learning \cite{bowlesFindingBenchmarks2024}. To analyze the memory--nonlinearity trade-off in QRC, we compute not only the IPC but also task-specific performance. 
We train our reservoir to predict the dynamics of the chaotic Mackey--Glass system \cite{MAC77}, a widely used benchmark in reservoir computing \cite{WER19, JAU24, TAN19c, APP11, BRU13a, LUE26, FUJ17, JAE04}.
We also consider the Lorenz task, in which the Lorenz time series is predicted, as it requires high nonlinear processing capabilities from the system \cite{LOR63, FUM25}. We choose these tasks specifically since their requirements for memory retention and nonlinear processing abilities are well-known, thereby enabling us to analyze and confirm the effect of the memory nonlinearity trade-off on the reservoir's task performance. \vspace{2mm}

\paragraph*{Lorenz.} The Lorenz 63 system is a standard benchmark for nonlinear time-series prediction~\cite{LOR63}. 
Its dynamics are described by the differential equations:

\begin{equation}
\begin{aligned}
\dot{x} &= \sigma (y - x), \\
\dot{y} &= x(\rho- z) - y, \\
\dot{z} &= xy - \beta z,
\end{aligned}
\label{eq:lorenz}
\end{equation}
with parameters $\sigma=10$, $\rho=28$, and $\beta = 8/3$. 
For numerical experiments, we integrate the system with $dt = 0.001$ and then discretize it with $\Delta t = 0.1$ as $x_n = x(n \Delta t)$, $y_n = y(n \Delta t)$, and $z_n = z(n \Delta t)$. Two benchmark tasks are considered: 
the \textit{Lorenz x--x} task (LXX), where $x_n$ is used to predict $x_{n+1}$, and the \textit{Lorenz x--z} task (LXZ), where $x_n$ is used to predict $z_n$. \vspace{2mm}

\paragraph*{Mackey--Glass.} The dynamic of the Mackey--Glass system 

\begin{equation}
\frac{dx}{dt} = \beta \frac{x(t - \tau_{MG})}{1 + x(t - \tau_{MG})^{10}} - \gamma x(t),
\end{equation}
exhibits chaotic behavior for $\tau_{MG} > 17$. Here, we choose $\beta = 0.2$, $\gamma = 0.1$, and $\tau_{MG} = 18$.
The time series is integrated using a time step of $dt = 0.1$ and then discretized with a sampling interval of $\Delta t = 3$ resulting in the discrete sequence $x_n = x(n \Delta t)$. 
The task is to predict $x_{n+1}$ given $x_n$.
A larger $\Delta t = 3$ is used since predicting further into the future increases the task's requirement for nonlinearity, enabling a more detailed test of the memory--nonlinearity trade-off.

%%%%%%%%%%%%%%%%%%%%%%%%%%%%%%   Memory-restrictiong %%%%%%%%%%%%%%%%%%%%%%%%%%%%%%%%%%
%%%%%%%%%%%%%%%%%%%%%%%%%%%%%%%%%%%%%%%%%%%%%%%%%%%%%%%%%%%%%%%%%%%%%%%%%%%%%%

\section{Memory-Restricted Quantum Reservoir Computing}\label{sec:restricted}
A central challenge in quantum reservoir computing remains the issue of measurement and the resulting collapse of the quantum state. In quantum mechanics, measuring a state $\ket{\phi}$ with respect to an observable $O$, defined by the eigenvalue equation $O\ket{o_i} = o_i\ket{o_i}$, induces the transition
\begin{align}
\ket{\phi} \xrightarrow{\mathrm{measurement}} \ket{o_i}, \quad p_i = \abs{\braket{o_i|\phi}}^2,
\end{align}
where the system collapses to the eigenstate $\ket{o_i}$ with probability $p_i$. This process inherently destroys information about the original quantum state, posing a fundamental limitation for reservoir-based computations.

In the first quantum reservoir computing scheme proposed by the authors of \cite{FUJ17}, the state matrix is constructed through successive measurements of the evolved system. This approach requires reinitializing the system with the complete input time series up to the current time step. Consequently, at each time step $t_n$, the state before measurement must be reconstructed, all inputs must be re-encoded, and the reservoir must be re-evolved. Given an input series of length $M$, this procedure results in a time complexity of
\begin{align}
    T_1(M) = \sum_{i=1}^{M} i = \frac{M(M+1)}{2} \in \mathcal{O}(M^2).
\end{align}
This quadratic scaling renders the method impractical for very long or continuous input sequences ($M \rightarrow \infty$).  

To address the time complexity in time-series tasks, weak measurement schemes \cite{MUJ23, FRA24} and reinitialization protocols \cite{CIN24} have been proposed. Other approaches include reintroducing measured outputs \cite{KOB24, AHM24} and continuous measurement schemes \cite{ZHU25}.  
Here, we briefly revisit the reinitialization scheme proposed in \cite{CIN24}, which was shown to not only reach the QRC limit but also enhance task performance and increase expressivity. Moreover, this approach reveals a clear memory–nonlinearity trade-off.
In this scheme, one considers at each time step the last $r$ inputs, given by
\begin{align}
    \mathbf{s}^{(n)} &= \{s_{n-r-1}, \ldots, s_{n-1}, s_n\} \nonumber \\
    &= \{s_1^{(n)}, s_2^{(n)}, \ldots, s_{r}^{(n)}\}.
\end{align}
For each input $s_n$, the system is first initialized to $\rho_0 = \ket{0000}\bra{0000}$.  
We first perform
\begin{align}
    \rho_1^{(\mathrm{in})} &= \sqrt{\frac{1 - {s}_1^{(n)}}{2}}\ket{0} + \sqrt{\frac{1 + {s}_1^{(n)}}{2}}\ket{1}, \nonumber \\
    \rho_1^{(n)} &= U_R\Big(\rho_1^{(\mathrm{in})} \otimes \Tr_1[\rho_0]\Big)U_R^\dagger.
\end{align}
For $k = 2, 3, \ldots, r$, this process is repeated iteratively as
\begin{align}
    \rho_k^{(\mathrm{in})} &= \sqrt{\frac{1 - {s}_k^{(n)}}{2}}\ket{0} + \sqrt{\frac{1 + {s}_k^{(n)}}{2}}\ket{1}, \nonumber \\
    \rho_k^{(n)} &= U_R\Big(\rho_k^{(\mathrm{in})} \otimes \Tr_1[\rho_{k-1}^{(n)}]\Big )U_R^{\dagger}.
\end{align}
At the end of the iteration, the observables are measured in the same manner as in the vanilla QRC algorithm.  
This approach requires only $r$ encodings and reservoir evolutions per time step, and therefore, for a time series of length $M$, the time complexity reduces to
\begin{align}
    T_2(M) = r M \in \mathcal{O}(M).
\end{align}

\begin{figure}[t]
  \centering
  \includegraphics[scale=1]{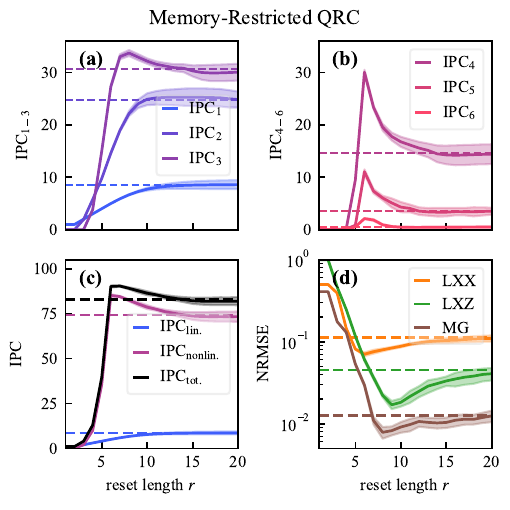}
  \caption{Dependence of the information processing capacities (IPCs) (a-c)  and task performance on the reset length $r$ (d) in the memory-restricted QRC.}
  \label{fig:LCQA_4q}
\end{figure}

\begin{table}[t]
\centering
\caption{Comparison between the vanilla QRC limits and the optimal results obtained with memory-restricted QRC (MRP). The best performance values and corresponding optimal reset length $r$ are listed for each metric.}
\label{tab:LCQA_results}
\renewcommand{\arraystretch}{1.3}
\setlength{\tabcolsep}{10pt}  % <-- increase column spacing
\begin{tabular}{lccc}
\toprule
 & \textbf{QRC} & \textbf{MRP} & { reset length $r$} \\
\midrule
$\mathrm{IPC}_1$ & 8.5 &  8.6 & 20 \\
$\mathrm{IPC}_2$ & 24.8 & 25.3 & 14 \\
$\mathrm{IPC}_3$ & 30.6 & 33.7 & 8 \\
$\mathrm{IPC}_4$ & 14.7 & 30.1 & 6 \\
$\mathrm{IPC}_5$ & 3.6 & 11.1 & 6 \\
$\mathrm{IPC}_6$ & 0.5 & 2.1 & 6 \\
$\mathrm{IPC}_{\mathrm{tot}}$ & 82.7 & 90.4 & 7 \\
\midrule
$\mathrm{LXX}$ & 0.111 & 0.0703 & 6 \\
$\mathrm{LXZ}$ & 0.045 & 0.017 & 9 \\
$\mathrm{MG}$ & 0.0126 & 0.0078 & 8 \\
\bottomrule
\end{tabular}
\end{table}
\Cref{fig:LCQA_4q} and \Cref{tab:LCQA_results} summarize the effect of the reset length $r$ on the information processing capacities (IPCs, blue to red) and the NRMSE, calculated using \Cref{eq:nrmse}, for the memory-restricted QRC.
\cref{fig:LCQA_4q}(a) shows the first three IPCs, which increase in dependence of the reset length and then saturate. $\mathrm{IPC}_1$ remains constant at approximately $8.5$ for $r>10$, indicating that the short-term memory capacity of the reservoir is preserved for larger reset lengths. It is also noteworthy that the vanilla QRC-limit is reached at this point, which is indicated by the horizontal line. The second $\mathrm{IPC}_2$ exhibits almost identical behavior to $\mathrm{IPC}_1$ and is only slightly larger than the vanilla QRC approach. The third order $\mathrm{IPC}_3$ however reaches a visibly larger value, which is indicated by the bold line being larger than the dashed line. \\
The higher-order IPCs displayed in \cref{fig:LCQA_4q}(b) exhibit a more pronounced dependence on the reset length. Both $\mathrm{IPC}_4$ and $\mathrm{IPC}_5$ increase sharply up to $r = 6$, reaching values more than twice those of the vanilla QRC limit (see \cref{tab:LCQA_results}). $\mathrm{IPC}_6$ also increases significantly, though its contribution remains smaller in magnitude. 
For all orders of the IPC, we observe the vanilla QRC limit reached for $r>15$, showcasing that the reservoir exhibits the fading memory property. 
The linear term, the nonlinear terms, and the total information processing capacity $\mathrm{IPC}_{\mathrm{tot}}$ are shown in \cref{fig:LCQA_4q}(c) in blue, purple, and black, respectively. We observe a peak of total IPC at $r = 6$ with a value of $90.5$, compared to $82.8$ for the vanilla QRC. For large reset lengths, the performance converges to that of the vanilla QRC, as indicated by the overlap between the bold and dashed lines. 
 
A restricted protocol that encodes only a few inputs enables the system to reconstruct these inputs more accurately \cite{CIN24}. At the same time, neither the Hamiltonian nor the unitary evolution change thus retaining the richness of the system dynamics provided by the quantum reservoir \cite{CIN25, CIN25a}. Reducing the number of encoded inputs while retaining the degrees of freedom of the time evolution results in these inputs being mapped in a more complex manner in Hilbert space, thereby enabling significantly increased nonlinearities. We also note, that the weights for the Hamiltonian considered here are sampled from $J_{ij} \in \mathcal{U}([0,1])$, after which the Hamiltonian is normalized 
to have spectral radius $\rho(H)=1$. In contrast, \cite{CIN24} sampled $J_{ij} \in \mathcal{U}([0.25, 0.75])$, resulting in different dynamics.

Finally, the task performances in \cref{fig:LCQA_4q}(d) show that all three benchmark tasks benefit from moderate reset lengths. The Lorenz $x \rightarrow x$ prediction improves from $\mathrm{NRMSE} = 0.111$ to $0.06$, the Lorenz $x \rightarrow z$ cross-prediction from $0.044$ to $0.015$, and the Mackey–Glass task from $0.011$ to $0.006$, with optimal performance typically obtained in the range $r \in [6,9]$. These trends align with the IPC analysis, confirming that moderate reset lengths enhance nonlinear processing and overall predictive accuracy without significantly compromising memory capacity.

%%%%%%%%%%%%%%%%%%%%%%% WEAK MEASUREMENT %%%%%%%%%%%%%%%%%%%%%%%%%%%%%%%%%%%
%%%%%%%%%%%%%%%%%%%%%%%%%%%%%%%%%%%%%%%%%%%%%%%%%%%%%%%%%%%%%%%%%%%%%%%%%%%%%%

\section{Quantum Reservoir Computing with Weak Measurements}\label{sec:weak}
Another possible strategy to address the time-complexity problem of the fully restarting protocol is the use of weak measurements for information extraction \cite{MUJ23}. Weak measurements induce less back-action on the system and therefore avoid restarting the reservoir. This advantage, however, comes at the cost of reduced information gain per measurement. Consequently, a larger number of individual trajectories must be evaluated to achieve noise levels comparable to strong measurements. Therefore a trade-off arises between the measurement strength and the runtime of the scheme, which depends on the specific hardware implementation. 

Beyond improvements in time efficiency, weak measurements have been observed to enhance task performance \cite{FRA24}. In this work, we argue that this effect stems from the memory--nonlinearity trade-off inherent to QRC. Additionally, we show that the measurement strength can serve as a tunable hyperparameter to adjust the memory or nonlinearity of the reservoir.
%this is due to the additional dissipation into the system introduced by the measurements, leading to a shift in the balance between memory retention and nonlinear processing abilities, which is tuneable via the measurement strength.

\begin{figure}
\centering
\includegraphics[width=0.8\linewidth]{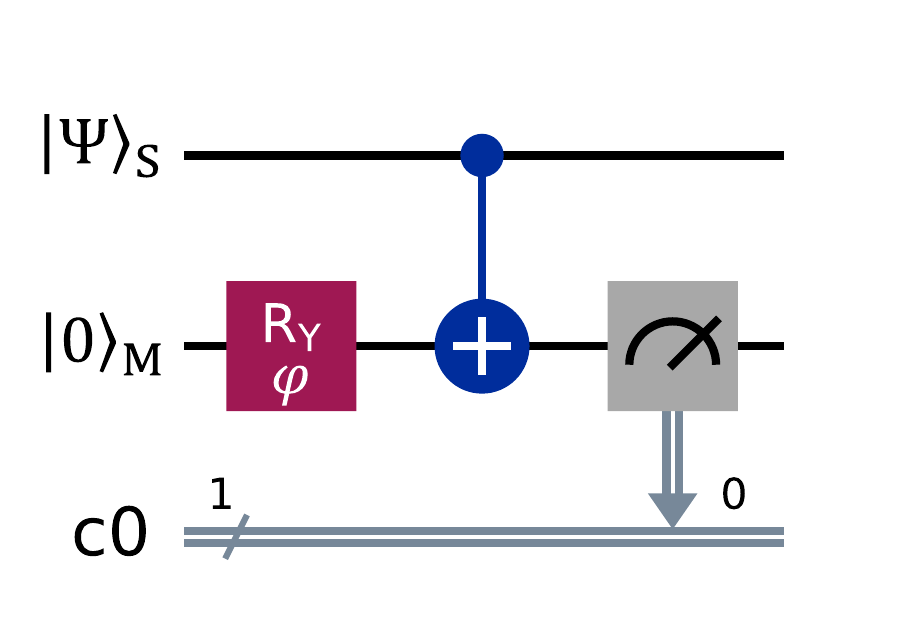}
\caption{Overview of the indirect measurement protocol. First, an ancilla qubit is initialized via an $R_Y (\varphi)$ gate, where the rotation angle $\varphi=\frac{\pi}{2}-\vartheta$ allows for tuning of the measurement strength $\vartheta$. This ancilla then interacts with the system qubit through a CNOT gate. Finally, a projective measurement is performed on the ancilla; the resulting outcome is used to infer the state of the system qubit.}
\label{fig:wmcircuit}
\end{figure}

To analyze this in detail, we utilize the previously introduced TFIM as the quantum reservoir. We further propose a weak-measurement protocol based on an ancilla-assisted measurement scheme \cite{brunWMTestonQCs2008}, as illustrated in \cref{fig:wmcircuit}. We motivate this protocol by noting that it can be directly implemented on current quantum hardware, where ongoing research could benefit from it \cite{chenQRConNISQ2020, KAW26}. In contrast, previous studies considered cavity-mediated indirect measurements \cite{MUJ23}.

In this scheme, tuning the rotation angle $\varphi = \frac{\pi}{2}-\vartheta$ allows for the control of the strength of the measurement $\vartheta$. The influence of the ancilla preparation on the subsequent CNOT gate provides an intuitive understanding of the mechanism. For the strong-measurement limit ($\vartheta=\frac{\pi}{2}$), the ancilla qubit remains in the $\ket{0}$ state, which maximizes entanglement with the system qubit through the CNOT gate. Consequently, a projective measurement of the ancilla effectively performs a projective measurement on the system. Conversely, in the weak-measurement limit ($\vartheta \to 0$), the ancilla qubit is rotated into an equal superposition $\frac{1}{\sqrt{2}}(\ket{0}+\ket{1})$, leaving the system and ancilla in a product state after the CNOT gate. In this case, no back-action is applied, but no information is obtained about the system's state either.
%We introduce this quantum circuit measurement scheme for QRC in contrast to other works that employ cavity-mediated indirect measurements \cite{MUJ23}, since it is readily applicable on currently existing quantum computers that can be exploited for quantum reservoir computing \cite{chenQRConNISQ2020}. 
To describe the effect of the measurement process on the reservoir, we model the back-action in an ensemble picture following \cite{MUJ23} as
\begin{align}
\rho_{k+1} = M \odot \left(U_R (\rho_k^\text{in} \otimes \Tr_1[\rho_k]) U_R^\dagger \right).
\end{align}
The effect of the measurement strength $\vartheta$ is encapsulated in the matrix $M$. Here, $\odot$ again denotes the element-wise product and $M$ is defined as
\begin{align} \label{eq:m_matrix}
M = \tilde{M}^{\otimes N} \thickspace \thickspace \thickspace \text{with} \thickspace \thickspace \thickspace 
\tilde{M} = \begin{pmatrix}
1 & \cos{\vartheta} \\
\cos{\vartheta}& 1 
\end{pmatrix}.
\end{align}
In this representation, the aforementioned limits of $\vartheta$ are readily apparent: $\vartheta=0$ corresponds to an infinitely weak measurement with no back-action ($M$ becomes a matrix of all ones), while $\vartheta=\frac{\pi}{2}$ recovers a projective measurement in the z-basis. To account for shot noise and reduced information gain of weaker measurements, the expectation value of the measured observable is sampled from a Gaussian distribution
\begin{align}
\langle \sz^{(i)}\rangle \sim \mathcal{N}\left(\Tr(\sz^{(i)}\rho), \left(\frac{1}{\sin(\vartheta)} \sqrt{\frac{1}{N_{\text{meas}}}}\right)^2\right),
\end{align} 
where its variance is inversely proportional to both the number of measurements $N_{\text{meas}}$ and the measurement strength $\vartheta$. In the strong measurement limit $(\vartheta=\pi/2)$ we recover the shot-noise scaling of equation \eqref{eq:shot_noise}, while a weaker measurement requires more shots to obtain an comparable noise level.

%All results shown are for a shot noise level of $10^{-5}$, where it should be noted that for weaker measurements more measurements have to be performed to reach the same noise level.
 
\begin{table}[t]
\centering
\caption{Comparison between the vanilla QRC limits and the optimal results obtained with the weak measurement protocol (WMP). The best performance values and corresponding optimal measurement strength $\vartheta$ are listed for each metric.}
\label{tab:WMQRC}
\renewcommand{\arraystretch}{1.3}
\setlength{\tabcolsep}{10pt}  % <-- increase column spacing
\begin{tabular}{lccc}
\toprule
 & \textbf{QRC} & \textbf{WMP} & { $\vartheta$} \\
\midrule
$\mathrm{IPC}_1$ & 8.5 &  8.5 & 0 \\
$\mathrm{IPC}_2$ & 24.8 & 24.8 & 0.021 \\
$\mathrm{IPC}_3$ & 30.6 & 33.1 & 0.109 \\
$\mathrm{IPC}_4$ & 14.7 & 20.1 & 0.152 \\
$\mathrm{IPC}_5$ & 3.6 & 4.6 & 0.131 \\
$\mathrm{IPC}_6$ & 0.5 & 1 & 0.152 \\
$\mathrm{IPC}_{\mathrm{tot}}$ & 82.7 & 87.7 & 0.109 \\
\midrule
$\mathrm{LXX}$ & 0.111 & 0.092 & 0.109 \\
$\mathrm{LXZ}$ & 0.045 & 0.026 & 0.109 \\
$\mathrm{MG}$ & 0.0126 & 0.0125 & 0.021 \\
\bottomrule
\end{tabular}
\end{table}
\begin{figure}[t]
  \centering
  \includegraphics[scale = 1]{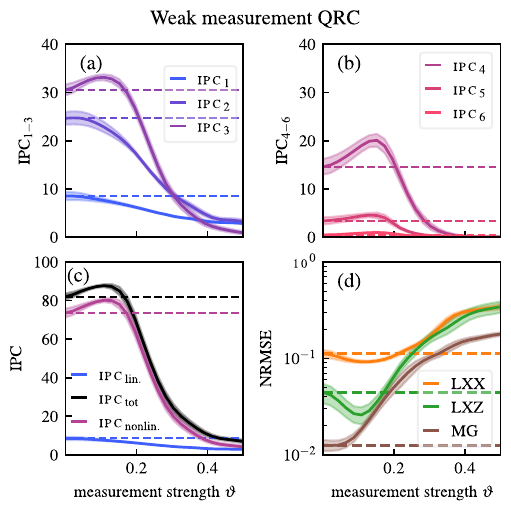}
  \caption{Dependence of the information processing capacities (IPCs) (a-c) and task performances (d) on the measurement strength $\vartheta$ defined in equation \eqref{eq:m_matrix}, where the dashed lines are indicating the vanilla QRC limit.}
  \label{fig:WMQRC}
\end{figure}

The results for varying measurement strengths $\vartheta$ in the weak measurement protocol are summarized in \Cref{tab:WMQRC} and \cref{fig:WMQRC}.  Panel (a) shows the first three degrees of the IPC. While $\text{IPC}_1$ and $\text{IPC}_2$ decrease for higher measurement strengths, the third degree $\text{IPC}_3$ initially increases from 30.6 to 33.1 before eventually decaying. This highlights the role of back-action: it degrades the system's memory retention, thereby lowering the memory-dependent first two IPC degrees, while simultaneously enhancing its nonlinear processing capacities. This shift in the memory-linearity trade-off is also evident in \cref{fig:WMQRC}(b), where $\text{IPC}_{4-6}$ exhibit a similar characteristic peak at moderate back-action. These results show that for all practical applications, the measurement strength must remain sufficiently weak; strong back-action suppresses all memory retention, rendering information processing impossible.

Tuning this trade-off via the measurement strength enables optimization of memory or nonlinearity. The overall processing capacity is maximized at $\vartheta = 0.109$, where $\text{IPC}_{\mathrm{tot}} = 87.7$, a $5\%$ improvement over the vanilla QRC limit of $82.7$ (see \cref{fig:WMQRC}(c) and \Cref{tab:WMQRC}). These findings are confirmed in \cref{fig:WMQRC}(d), which shows the task-specific performance of the WMP. For the Lorenz $x \rightarrow x$ and $x \rightarrow z$ prediction the NRMSE decreases (from 0.111 to 0.092 and 0.045 to 0.026, respectively) for a moderate amount of back-action. This confirms our earlier findings that a moderate amount of back-action increases the nonlinear processing capacities, therefore improving the performance of tasks that require nonlinear processing. Conversely, the Mackey-Glass task, which necessitates long-term memory, performs best at a very weak measurement strength of $\vartheta=0.021$, where memory retention is maximal.

In summary the weak measurement protocol is particularly suited for nonlinear processing tasks. Tuning the measurement strength $\vartheta$ allows the exploitation of the memory-nonlinearity trade-off, maximizing both the performance and reducing the number of measurements required to obtain a certain noise level. The WMP therefore also offers an intuitive way to study and exploit the memory-nonlinearity trade-off and its influence on task-specific performance in QRC.

%%%%%%%%%%%%%%%%%%%%%%%%%%  CHAOS  %%%%%%%%%%%%%%%%%%%%%%%%%%%%%%%%%%
%%%%%%%%%%%%%%%%%%%%%%%%%%%%%%%%%%%%%%%%%%%%%%%%%%%%%%%%%%%%%%%%%%%%%%%%%%%%%%
\section{Quantum Reservoir Computing at the Edge of Chaos}\label{sec:chaos}
One major aspect is the choice of the Hamiltonian that governs the reservoir dynamics. Although this behavior is often determined by the underlying physical system, different Hamiltonians can exhibit qualitatively distinct dynamics, and phase transitions can play a significant role \cite{MAR21}. In that work, the authors analyzed the transverse-field Ising model with on-site disorder in both ergodic and localized regimes and demonstrated that optimal task performance occurs near the boundary between these phases, at the so-called \textit{edge of quantum chaos}.
More recent work by Kobayashi and Motome \cite{KOB25} extended this line of investigation to the Sachdev Ye Kitaev (SYK) model, showing that the transition between chaotic and non chaotic phases strongly influences the computational capability of the reservoir. Their results suggest that QRC achieves maximal expressivity and task performance in the vicinity of a phase transition, when the system maintains a balance between memory retention and nonlinear response. 

We again consider the Ising Hamiltonian described by \cref{eq:ising_chaos}, but sample the coupling coefficients independently and uniformly as \(J_{ij} \sim \mathcal{U}([-1,1])\), without applying any normalization.
To capture statistical behavior, we generate one hundred distinct Hamiltonians and systematically vary the external field \( h \). Around \( h \approx 0.5 \), the system undergoes a transition between a regular and a chaotic phase. For \( h > 0.5 \), the dynamics become increasingly chaotic, leading to enhanced nonlinearity and reduced memory, which are characteristic features of the quantum edge of chaos regime.

\begin{figure}[t]
  \centering
  \includegraphics[scale=1]{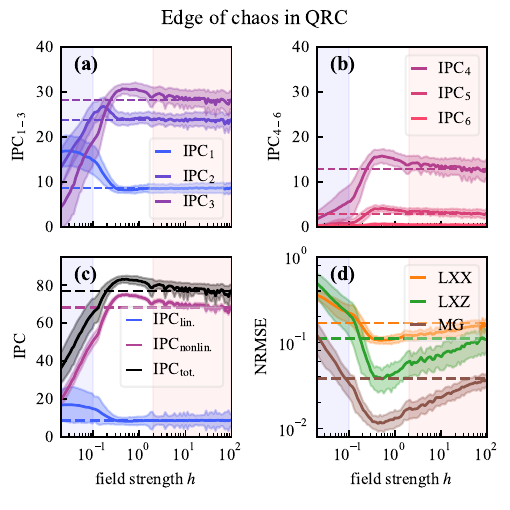}
  \caption{Dependence of the information processing capacities (IPCs) (a-c) and task performance (d) on the field strength $h$ of the Hamiltonian (\cref{eq:ising_chaos}). Averages are computed over 100 sampled Hamiltonians. Blue and red region represent the localized and ergodic regime respectively. The optimal regime appears near the transition point $h \approx 0.5$, corresponding to the onset of quantum chaos.}
  \label{fig:chaos_4q}
\end{figure}

\begin{table}[t]
\centering
\caption{Best performance values and corresponding optimal external field strength $h$ of the TFIM Hamiltonian (\cref{eq:ising_chaos}).}
\label{tab:chaos_results}
\renewcommand{\arraystretch}{1.3}
\setlength{\tabcolsep}{10pt}  % <-- increase column spacing
\begin{tabular}{lccc}
\toprule
& ergodic & Best  & $h$  \\
\midrule
$\mathrm{IPC}_1$ & 8.7 & 16.7 &  0.03  \\
$\mathrm{IPC}_2$ & 23.8 & 26.8 &  0.19 \\
$\mathrm{IPC}_3$ & 28.7 & 30.7 & 0.43 \\
$\mathrm{IPC}_4$ & 12.9 & 15.8 & 0.52 \\
$\mathrm{IPC}_5$ & 2.9 & 4.1 & 0.52 \\
$\mathrm{IPC}_6$ & 0.5 & 0.6 & 0.358 \\
$\mathrm{IPC}_{\mathrm{tot}}$ &76.9 & 83.2 & 0.430  \\
\midrule
$\mathrm{LXX}$ & 0.1700 & 0.1063 & 0.472  \\
$\mathrm{LXZ}$ & 0.1112 & 0.0375 & 0.520  \\
$\mathrm{MG}$ & 0.0382 &  0.0114 & 0.430 \\
\bottomrule
\end{tabular}
\end{table}
\Cref{fig:chaos_4q} and \cref{tab:chaos_results} illustrate how the external field \( h \) affects the information processing capacities (IPCs) and task performance. 
\Cref{fig:chaos_4q}(a,b) shows that the IPCs increase visibly as \(h\) approaches the transition region around \(h \approx 0.3\text{--}0.5\). In the weak-field regime ($h \leq 0.1$, indicated by the blue background), the system is in a localized phase, resulting in low nonlinearity. However, these simple dynamics favor memory retention, as reflected in the large short-term memory capacity $\mathrm{IPC}_1$, which reaches a maximum of $16.7$ in this regime before dropping to $8.6$ in the chaotic phase (red background) for $h>2$. $\mathrm{IPC}_2$ peaks at $h = 0.19$, well below the transition, while $\mathrm{IPC}_3$ attains its maximum at $h = 0.43$. Higher-order terms such as $\mathrm{IPC}_5$ and $\mathrm{IPC}_6$ peak near the transition, indicating enhanced nonlinearity.
The memory capacity $\mathrm{IPC}_{\mathrm{lin.}}$ (blue), nonlinear capacities $\mathrm{IPC}_{\mathrm{nonlin.}}$ (purple), and the total information processing capacity $\mathrm{IPC}_{\mathrm{tot.}}$ (black) follow a similar pattern, as shown in \cref{fig:chaos_4q}(c). A maximum of $\mathrm{IPC}_{\mathrm{tot}} = 83.2$ is reached at $h = 0.43$. This region marks a balance between memory and nonlinearity and is often associated with the ``edge of quantum chaos,'' where the reservoir attains maximal expressivity and computational richness.\\
As \( h \) increases toward the chaotic phase, all higher order IPCs rise significantly, reaching their maxima near the transition, as summarized in \cref{tab:chaos_results}. Beyond this point, the capacities saturate or decline slightly, indicating that excessive chaos reduces effective memory.  

The task performance curves in \cref{fig:chaos_4q}(d) support this interpretation. All benchmark tasks achieve minimal prediction error in the region between the localized phase (blue) and the ergodic phase (red). The Lorenz \( x \rightarrow x \) task reaches \(\mathrm{NRMSE} = 0.106\), the Lorenz \( x \rightarrow z \) task \( 0.037 \), and the Mackey Glass task \( 0.008 \). For stronger fields (\( h > 2 \)), performance worsens, consistent with the decreasing nonlinear IPCs.

Overall, these results show that the optimal computational regime of quantum reservoir computing (QRC) emerges near the phase transition between regular and chaotic dynamics, driven by the intrinsic trade-off between memory and nonlinearity observed at the transition. %Additionally, this raises the question of whether a quantum-inspired topology based on these results can serve as a tool to design effective topologies in classical reservoir computing.

%%%%%%%%%%%%%%%%%%%%% DISSIPATION LINDBLAD    %%%%%%%%%%%%%%%%%%%%%%%%%%
%%%%%%%%%%%%%%%%%%%%%%%%%%%%%%%%%%%%%%%%%%%%%%%%%%%%%%%%%%%%%%%%%%%%%%%%%%%%%%
\section{Dissipative Quantum Reservoir Computing}\label{sec:lindblad}
In the above QRC implementations, the fading memory necessary for computation is induced by the dissipative input mechanism.
Experimental realizations of quantum reservoir computers will always be subject to interactions with the environment, resulting in additional dissipative dynamics causing memory losses.
While these effects are unwanted in gate-based quantum computers, we here exploit them as a resource for reservoir computing \cite{SAN24,OLI23}.
By combining the dissipative dynamics of open quantum systems with a unitary input encoding instead of the previous state-reset approaches, we obtain a QRC implementation that features a natural form of fading memory without a lower bound.

We model the dynamical quantum system serving as reservoir with a coherently controlled TFIM.
The control is amplitude modulated according to the input signal strength $s(t)$ and introduced to the model via an additional $\sy$ term on the first qubit in the TFIM Hamiltonian.
The first qubit is chosen as input in accordance with the above QRC schemes -- in principle, any number of qubits may be subjected to the input signal.
In an experimental setup, this may be implemented by a continuous-wave laser, encoding the input.
We express the resulting controlled-TFIM Hamiltonian in the rotating frame of the laser beam as
\begin{equation}
  H(t) = \sum_{i<j} J_{ij} \left(\splus^{(i)} \sm^{(j)} + \sm^{(i)} \splus^{(j)}\right) + s(t)\sy^{(1)}.
  \label{eq:tfim_td}
\end{equation}
with $\splus^{(i)}$ and $\sm^{(i)}$ being the raising and lowering operators for qubit $i$, respectively.

While this model provides a dynamical quantum system with an input channel, it is not suitable for reservoir computing, due to its strictly unitary evolution, as shown in \cite{GOE25}.
However, the aforementioned dissipative dynamics lifts this barrier, providing a fully functional QRC implementation.
For simplicity, we here model the dissipation as a constant qubit decay $\gamma$ applied globally to all qubits.
This yields the Lindblad master equation for the full dynamics:
\begin{equation}
  \dot{\rho} = -\ii [H(t), \rho] + \gamma \sum_i \left(\sm^{(i)}\rho \splus^{(i)} - \frac{1}{2}\left\{\splus^{(i)}\sm^{(i)}, \rho\right\}\right),
  \label{eq:lindblad}
\end{equation}

\begin{figure}[t]
  \centering
  \includegraphics[scale=1]{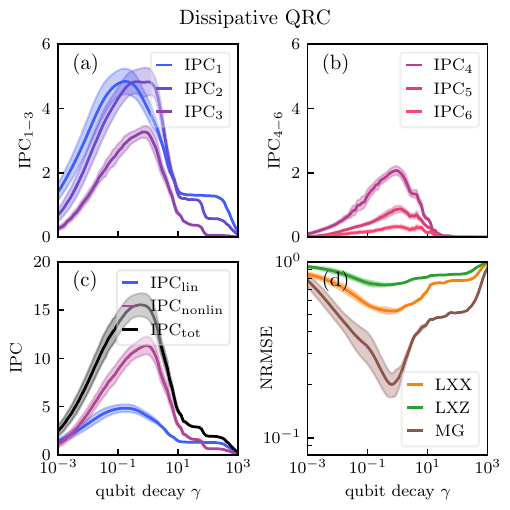}
  \caption{Dependence of the information processing capacities (IPCs) and task performance on the qubit decay $\gamma$ defined in \cref{eq:lindblad}.}
  \label{fig:lindblad_4q}
\end{figure}

Due to the increased computational complexity of this dynamical approach, we set a shorter input injection period of $T = 1$ and a lower time multiplexing of $N_V = 10$ steps, compared to the state-reset approaches.
These restrictions result in overall worse performance compared to the other reservoir setups, as the information does not have enough time to completely scramble throughout the system \cite{DOM24, CIN25, CIN25a, VET25, SHE20,MI21,LOM24}.
The general trend of the memory-nonlinearity trade-off, is, however, not affected by these limitations, further underlining its universality.

In Fig.~\ref{fig:lindblad_4q}, the IPCs, the Lorenz task performance, and the Mackey-Glass task performance are shown for a four-qubit dynamical TFIM.
In this implementation, the parameter controlling the memory-nonlinearity trade-off is the qubit decay $\gamma$, which is varied along the x-axes.
Due to the purely unitary model featuring no fading-memory at all, the performance metrics have no finite asymptotic values for very large or small $\gamma$, but instead vanish in both regimes.
This allows more flexibility than the standard state-reset approach, which has a fixed minimum decoherence rate imposed by the input mechanism.

In tuning the dissipation strength $\gamma$, we find sweet-spot regimes for the linear STMC as well as the nonlinear IPCs.
Panels a) and b) of Fig.~\ref{fig:lindblad_4q} display the performance of each individual IPC and are summarized in panel c), where the total IPC is shown in comparison to the linear and nonlinear capacities.
The general sweet spot of the IPCs coincides with the findings in \cite{GOE25}, where a similar input channel was employed.
Importantly, we observe the sweet-spot for nonlinear IPCs to only emerge at larger dissipation rates, that already exhibit a decrease of linear memory.
Thus, also in this QRC implementation, the memory-nonlinearity trade-off that is central to this work materializes, which forbids maximal linear memory to appear at the same system configuration as the maximal nonlinear processing powers.
As expected, the performance for any degree vanishes for too little ($\lesssim 10^{-2}$) or too much ($\gtrsim 10^2$) dissipation.

Both one-step-ahead prediction tasks -- the Lorenz $x\to x$-coordinate prediction and the Mackey-Glass prediction -- reach optimal performance around the peak of the total IPC.
As discussed for the previous QRC frameworks, this is a consequence of the rather large time steps in the data requiring more nonlinear processing capabilities.
The influence of memory becomes apparent in the Lorenz $x\to z$ cross-prediction.
Here, the coupling dynamics evolves over long periods of time, and memory retention becomes the paramount feature of the reservoir.
As nonlinear processing is still necessary for this task, the optimal Lorenz $x\to z$ performance does not exactly coincide with the linear IPC peak.

\begin{table}[h]
\centering
\caption{Optimal Lindblad QRC performance metrics for all analysed tasks with respective dissipation strength $\gamma$.}
\label{tab:lindblad_results}
\renewcommand{\arraystretch}{1.3}
\setlength{\tabcolsep}{10pt}  % <-- increase column spacing
\begin{tabular}{lccc}
\toprule
 & Best  & $\gamma$  \\
\midrule
$\mathrm{IPC}_1$ & 4.84 & 0.16 \\
$\mathrm{IPC}_2$ & 4.84 & 0.38 \\
$\mathrm{IPC}_3$ & 3.27 & 0.77 \\
$\mathrm{IPC}_4$ & 2.08 & 0.92 \\
$\mathrm{IPC}_5$ & 0.88 & 1.30 \\
$\mathrm{IPC}_6$ & 0.33 & 0.92 \\
$\mathrm{IPC}_{\mathrm{tot}}$ & 15.56 & 0.54 \\
\midrule
$\mathrm{LXX}$ & 0.52 & 0.77  \\
$\mathrm{LXZ}$ & 0.74 & 0.38  \\
$\mathrm{MG}$ & 0.20 & 0.65  \\
\bottomrule
\end{tabular}
\end{table}
While absolute performance metrics of the dissipative quantum reservoir are not directly comparable to other setups due to the shorter time scales in which information spreads in the system and sparser multiplexing, the dynamically driven approach has proven capable of processing complex information.
We summarize all results in \cref{tab:lindblad_results}.
The memory-nonlinearity trade-off is particularly pronounced when comparing the linear memory to the second order IPCs, whose peaks differ by one order of magnitude in the dissipation strength.
This underlines how nonlinear processing benefits from quickly contracting state evolution maps, whereas linear memory needs information to remain in the quantum system for longer times.
Importantly, the proposed setup is not vulnerable to additional dissipative effects in real-world scenarios, but benefits from and even relies on natural dissipation of moderate strength.
This makes the dynamically-driven, dissipative quantum reservoir computer a compelling candidate for hardware implementations of QRC.

\section{Conclusion} 

This work investigated the fundamental dynamical principles that govern the performance of quantum reservoir computing (QRC) across a wide range of architectures. While many QRC variants have been shown to yield improved performance in specific settings, the reasons behind these gains have remained fragmented and difficult to relate to one another. By employing the information processing capacity, a dynamical measure originating from nonlinear systems theory, we demonstrated that these seemingly unrelated approaches are in fact manifestations of the same underlying mechanism: a balance between memory retention and nonlinear response. Through a systematic analysis of memory constraints, weak measurements, phase transitions, and dissipative dynamics, we show that these protocols induce a tunable trade-off between memory and nonlinearity.

In the restricted protocol, the reset length controls how far back the reservoir can retain relevant information, and we showed that appropriately chosen reset lengths significantly enhance the nonlinear components of the information processing capacity and lead to improved task performance. In the weak measurement framework, the measurement strength determines how much information is extracted at each time step while governing the disturbance of the internal reservoir state. We found that intermediate measurement strengths optimize this balance, yielding both strong nonlinear processing and favorable computational performance. At the edge of quantum chaos, we observed an analogous effect. By tuning the transverse field in the quantum Ising model, we showed that reservoirs close to the localization–ergodicity transition simultaneously retain memory from past inputs and exhibit rich nonlinear dynamics, resulting in both large IPC values and superior performance on the benchmark tasks. Finally, in dissipative reservoirs governed by Lindblad dynamics, we identified an intermediate dissipation regime in which fading memory is optimal. This regime again maximizes the IPC and leads to the most accurate predictions, confirming that controlled dissipation provides a practical mechanism for tailoring the memory–nonlinearity trade-off on near-term hardware.

These findings provide a unified dynamical perspective on QRC and identify the memory–nonlinearity trade-off as a central design principle underlying quantum reservoir architectures. By framing diverse QRC approaches within a common theoretical language, our work offers both conceptual clarity and practical guidance for the design of quantum reservoirs. We expect that this framework will be useful for analyzing more advanced architectures, guiding hardware implementation strategies, and ultimately providing insight into how quantum systems can be engineered to process temporal information most efficiently.

\begin{acknowledgments}
    N.G., L.G., and C.G. thank Frederik Lohof for numerous helpful discussions.
    N.G. and C.G. gratefully acknowledge funding by the Deutsche Forschungsgemeinschaft (German Science Foundation, DFG) via the project PhotonicQRC (Gi1121/6-1). L.G. acknowledges funding by the Deutsche Forschungsgemeinschaft (DFG, German Research Foundation) via the Research Unit FOR 5688 (Project No. 521530974).
\end{acknowledgments}

\bibliography{lit}

@aricle{OLI23,
  title = {Benefits of {{Open Quantum Systems}} for {{Quantum Machine Learning}}},
  author = {{Olivera-Atencio}, Mar{\'i}a Laura and Lamata, Lucas and {Casado-Pascual}, Jes{\'u}s},
  year = {2023},
  journal = {Advanced Quantum Technologies},
  pages = {2300247},
  issn = {2511-9044},
  doi = {10.1002/qute.202300247},
  urldate = {2023-12-15},
}

@article{SHE20,
  title = {Information {{Scrambling}} in {{Quantum Neural Networks}}},
  author = {Shen, Huitao and Zhang, Pengfei and You, Yi-Zhuang and Zhai, Hui},
  year = {2020},
  month = may,
  journal = {Physical Review Letters},
  volume = {124},
  number = {20},
  pages = {200504},
  publisher = {American Physical Society},
  doi = {10.1103/PhysRevLett.124.200504},
}

@article{LOM24,
  title = {An Operational Definition of Quantum Information Scrambling},
  author = {Lo Monaco, Gabriele and Innocenti, Luca and Cilluffo, Dario and Chisholm, Diana A and Lorenzo, Salvatore and Massimo Palma, G},
  year = {2024},
  month = dec,
  journal = {Quantum Science and Technology},
  volume = {10},
  number = {1},
  pages = {015055},
  publisher = {IOP Publishing},
  issn = {2058-9565},
  doi = {10.1088/2058-9565/ad9ed2},
}

@article{MI21,
  title = {Information Scrambling in Quantum Circuits},
  author = {Mi, Xiao and Roushan, Pedram and Quintana, Chris and Mandr{\`a}, Salvatore and Marshall, Jeffrey and Neill, Charles and Arute, Frank and Arya, Kunal and Atalaya, Juan and Babbush, Ryan and Bardin, Joseph C. and Barends, Rami and Basso, Joao and Bengtsson, Andreas and Boixo, Sergio and Bourassa, Alexandre and Broughton, Michael and Buckley, Bob B. and Buell, David A. and Burkett, Brian and Bushnell, Nicholas and Chen, Zijun and Chiaro, Benjamin and Collins, Roberto and Courtney, William and Demura, Sean and Derk, Alan R. and Dunsworth, Andrew and Eppens, Daniel and Erickson, Catherine and Farhi, Edward and Fowler, Austin G. and Foxen, Brooks and Gidney, Craig and Giustina, Marissa and Gross, Jonathan A. and Harrigan, Matthew P. and Harrington, Sean D. and Hilton, Jeremy and Ho, Alan and Hong, Sabrina and Huang, Trent and Huggins, William J. and Ioffe, L. B. and Isakov, Sergei V. and Jeffrey, Evan and Jiang, Zhang and Jones, Cody and Kafri, Dvir and Kelly, Julian and Kim, Seon and Kitaev, Alexei and Klimov, Paul V. and Korotkov, Alexander N. and Kostritsa, Fedor and Landhuis, David and Laptev, Pavel and Lucero, Erik and Martin, Orion and McClean, Jarrod R. and McCourt, Trevor and McEwen, Matt and Megrant, Anthony and Miao, Kevin C. and Mohseni, Masoud and Montazeri, Shirin and Mruczkiewicz, Wojciech and Mutus, Josh and Naaman, Ofer and Neeley, Matthew and Newman, Michael and Niu, Murphy Yuezhen and O'Brien, Thomas E. and Opremcak, Alex and Ostby, Eric and Pato, Balint and Petukhov, Andre and Redd, Nicholas and Rubin, Nicholas C. and Sank, Daniel and Satzinger, Kevin J. and Shvarts, Vladimir and Strain, Doug and Szalay, Marco and Trevithick, Matthew D. and Villalonga, Benjamin and White, Theodore and Yao, Z. Jamie and Yeh, Ping and Zalcman, Adam and Neven, Hartmut and Aleiner, Igor and Kechedzhi, Kostyantyn and Smelyanskiy, Vadim and Chen, Yu},
  year = {2021},
  month = dec,
  journal = {Science},
  volume = {374},
  number = {6574},
  pages = {1479--1483},
  publisher = {American Association for the Advancement of Science},
  doi = {10.1126/science.abg5029},
}

@ARTICLE{NIE06a,
    author = {Nielsen, M. A. and Dowling, M. R. and Gu, M. and Doherty, A. C.},
     month = feb,
     title = {Quantum {Computation} as {Geometry}},
   journal = {Science},
    volume = {311},
    number = {5764},
      year = {2006},
     pages = {1133--1135},
      issn = {0036-8075, 1095-9203},
       url = {https://www.science.org/doi/10.1126/science.1121541},
       doi = {10.1126/science.1121541},
}

@ARTICLE{ARU19,
    author = {Arute, F. and Arya, K. and Babbush, R. and Bacon, D. and Bardin, J. C. and Barends, R. and Biswas, R. and Boixo, S. and Brandao, F. G. S. L. and Buell, D. A. and Burkett, B. and Chen, Y. and Chen, Z. and Chiaro, B. and Collins, R. and Courtney, W. and Dunsworth, A. and Farhi, E. and Foxen, B. and Fowler, A. and Gidney, C. and Giustina, M. and Graff, R. and Guerin, K. and Habegger, S. and Harrigan, M. P. and Hartmann, M. J. and Ho, A. and Hoffmann, M. and Huang, T. and Humble, T. S. and Isakov, S. V. and Jeffrey, E. and Zhang, Z. and Kafri, D. and Kechedzhi, K. and Kelly, J. and Klimov, P. V. and Knysh, S. and Korotkov, A. and Kostritsa, F. and Landhuis, D. and Lindmark, M. and Lucero, E. and Lyakh, D. and Mandr{\`{a}}, S. and McClean, J. R. and McEwen, M. and Megrant, A. and Mi, X. and Michielsen, K. and Mohseni, M. and Mutus, J. and Naaman, O. and Neeley, M. and Neill, C. and Niu, M. Y. and Ostby, E. and Petukhov, A. and Platt, J. C. and Quintana, C. and Rieffel, E. G. and Roushan, P. and Rubin, N. C. and Sank, D. and Satzinger, K. J. and Smelyanskiy, V. and Sung, K. J. and Trevithick, M. D. and Vainsencher, A. and Villalonga, B. and White, T. and Yao, Z. J. and Yeh, P. and Zalcman, A. and Neven, H. and Martinis, J. M.},
     month = {October},
     title = {Quantum supremacy using a programmable superconducting processor},
   journal = {Nature},
    volume = {574},
    number = {7779},
      year = {2019},
     pages = {505--510},
      issn = {1476-4687},
       url = {https://www.nature.com/articles/s41586-019-1666-5},
       doi = {10.1038/s41586-019-1666-5},
}

@ARTICLE{SCH15a,
    author = {Schuld, M. and Sinayskiy, I. and Petruccione, F.},
     month = apr,
     title = {An introduction to quantum machine learning},
   journal = {Contemp. Phys.},
    volume = {56},
    number = {2},
      year = {2015},
     pages = {172--185},
      note = {Publisher: Taylor \& Francis \_eprint: https://doi.org/10.1080/00107514.2014.964942},
      issn = {0010-7514},
       url = {https://doi.org/10.1080/00107514.2014.964942},
       doi = {10.1080/00107514.2014.964942},
}

@ARTICLE{BIA17,
    author = {Biamonte, J. and Wittek, P. and Pancotti, N. and Rebentrost, P. and Wiebe, N. and Lloyd, S.},
     month = {September},
     title = {Quantum machine learning},
   journal = {Nature},
    volume = {549},
    number = {7671},
      year = {2017},
     pages = {195--202},
      issn = {1476-4687},
       url = {https://www.nature.com/articles/nature23474},
       doi = {10.1038/nature23474},
}

@ARTICLE{CER22,
    author = {Cerezo, M. and Verdon, G. and Huang, H. and Cincio, L. and Coles, P. J.},
     month = {September},
     title = {Challenges and opportunities in quantum machine learning},
   journal = {Nat. Comput. Sci..},
    volume = {2},
    number = {9},
      year = {2022},
     pages = {567--576},
      issn = {2662-8457},
       url = {https://www.nature.com/articles/s43588-022-00311-3},
       doi = {10.1038/s43588-022-00311-3},
}

@ARTICLE{FUJ17,
    author = {Fujii, K. and Nakajima, K.},
     month = aug,
     title = {Harnessing {Disordered}-{Ensemble} {Quantum} {Dynamics} for {Machine} {Learning}},
   journal = {Phys. Rev. Applied},
    volume = {8},
    number = {2},
      year = {2017},
     pages = {024030},
      note = {Publisher: American Physical Society},
       url = {https://link.aps.org/doi/10.1103/PhysRevApplied.8.024030},
       doi = {10.1103/physrevapplied.8.024030},
}

@TECHREPORT{JAE01,
       author = {Jaeger, H.},
        title = {The ``echo state'' approach to analysing and training recurrent neural networks},
         type = {GMD Report},
       number = {148},
         year = {2001},
  institution = {GMD - German National Research Institute for Computer Science},
          doi = {10.24406/publica-fhg-291111}, 
}

@ARTICLE{SUZ22,
    author = {Suzuki, Y. and Gao, Q. and Pradel, K. C. and Yasuoka, K. and Yamamoto, N.},
     month = jan,
     title = {Natural quantum reservoir computing for temporal information processing},
   journal = {Sci. Rep.},
    volume = {12},
    number = {1},
      year = {2022},
     pages = {1353},
      note = {Publisher: Nature Publishing Group},
      issn = {2045-2322},
       url = {https://www.nature.com/articles/s41598-022-05061-w},
       doi = {10.1038/s41598-022-05061-w}, 
}

@ARTICLE{NAK19,
    author = {Nakajima, K. and Fujii, K. and Negoro, M. and Mitarai, K. and Kitagawa, M.},
     month = mar,
     title = {Boosting {Computational} {Power} through {Spatial} {Multiplexing} in {Quantum} {Reservoir} {Computing}},
   journal = {Phys. Rev. Applied},
    volume = {11},
    number = {3},
      year = {2019},
     pages = {034021},
      note = {Publisher: American Physical Society},
       url = {https://link.aps.org/doi/10.1103/PhysRevApplied.11.034021},
       doi = {10.1103/physrevapplied.11.034021}, 
}

@ARTICLE{GAR23,
    author = {Garci­a-Beni, J. and Giorgi, G. L. and Soriano, M. C. and Zambrini, R.},
     title = {Scalable photonic platform for real-time quantum reservoir computing},
   journal = {Phys. Rev. Applied},
    volume = {20},
      year = {2023},
     pages = {014051},
       doi = {10.1103/PhysRevApplied.20.014051}, 
}

@ARTICLE{MUJ21a,
    author = {Mujal, P. and Martinez-Pena, R. and Nokkala, J. and Garci­a-Beni, J. and Giorgi, G. L. and Soriano, M. C. and Zambrini, R.},
     title = {Opportunities in Quantum Reservoir Computing and Extreme Learning Machines},
   journal = {Adv. Quantum Technol.},
    volume = {4},
    number = {8},
      year = {2021},
     pages = {2100027},
       doi = {https://doi.org/10.1002/qute.202100027}, 
}

@ARTICLE{ABB24a,
    author = {Abbas, A. H. and Abdel-Ghani, H. and Maksymov, I. S.},
     title = {Classical and {Quantum} {Physical} {Reservoir} {Computing} for {Onboard} {Artificial} {Intelligence} {Systems}: {A} {Perspective}},
   journal = {Dynamics},
    volume = {4},
    number = {3},
      year = {2024},
     pages = {643-670},
      issn = {2673-8716},
       doi = {10.3390/dynamics4030033}
}

@INPROCEEDINGS{FUJ21,
     author = {Fujii, K. and Nakajima, K.},
     editor = {Nakajima, K. and Fischer, Ingo},
      title = {Quantum {Reservoir} {Computing}: {A} {Reservoir} {Approach} {Toward} {Quantum} {Machine} {Learning} on {Near}-{Term} {Quantum} {Devices}},
       year = {2021},
      pages = {423--450},
  publisher = {Springer Singapore},
    address = {Singapore},
       isbn = {9789811316869 9789811316876},
        url = {https://link.springer.com/10.1007/978-981-13-1687-6_18},
        doi = {10.1007/978-981-13-1687-6_18}, 
}

@ARTICLE{NAK20,
    author = {Nakajima, K.},
     title = {Physical reservoir computing—an introductory perspective},
   journal = {Jpn. J. Appl. Phys.},
    volume = {59},
    number = {6},
      year = {2020},
     pages = {060501},
       doi = {10.35848/1347-4065/ab8d4f}
}

@ARTICLE{PFE23,
    author = {Pfeffer, P. and Heyder, F. and Schumacher, J.},
     title = {Reduced-order modeling of two-dimensional turbulent Rayleigh-B{\'{e}}nard flow by hybrid quantum-classical reservoir computing},
   journal = {Phys. Rev. Res.},
    volume = {5},
    number = {4},
      year = {2023},
     pages = {043242},
       url = {https://link.aps.org/doi/10.1103/PhysRevResearch.5.043242},
       doi = {10.1103/PhysRevResearch.5.043242},
}

@ARTICLE{KAW26,
     author = {Kawanabe, M. and {\v C}indrak, S. and L{\"{u}}dge, K. and Shirakashi, J. and Shibuya, T. and Imai, H.},
      month = feb,
      title = {Efficient time-series prediction on {NISQ} devices via time-delayed quantum extreme learning machine},
    journal = {arXiv:2602.21544},
       year = {2026},
  publisher = {arXiv},
        url = {http://arxiv.org/abs/2602.21544}
}

@ARTICLE{GOE23,
    author = {G{\"{o}}tting, N. and Lohof, F. and Gies, C.},
     month = {November},
     title = {Exploring quantumness in quantum reservoir computing},
   journal = {Phys. Rev. A},
    volume = {108},
    number = {5},
      year = {2023},
     pages = {052427},
       url = {https://link.aps.org/doi/10.1103/PhysRevA.108.052427},
       doi = {10.1103/physreva.108.052427}, 
}

@ARTICLE{VET25,
    author = {Vetrano, M. and Lo Monaco, G. and Innocenti, L. and Lorenzo, S. and Palma, G. M.},
     month = {February},
     title = {State estimation with quantum extreme learning machines beyond the scrambling time},
   journal = {npj Quantum Inf},
    volume = {11},
    number = {1},
      year = {2025},
     pages = {1--8},
      issn = {2056-6387},
       url = {https://www.nature.com/articles/s41534-024-00927-5},
       doi = {10.1038/s41534-024-00927-5}, 
}

@article{XIO25,
	title = {On fundamental aspects of quantum extreme learning machines},
	volume = {7},
	issn = {2524-4914},
	url = {https://doi.org/10.1007/s42484-025-00239-7},
	doi = {10.1007/s42484-025-00239-7},
	number = {1}, 
	journal = {Quantum Mach. Intell.},
	author = {Xiong, W. and Facelli, G. and Sahebi, M. and Agnel, O. and Chotibut, T. and Thanasilp, S. and Holmes, Z.}, 
	year = {2025},
	pages = {20}, 
}

@ARTICLE{DOM24,
    author = {Domingo, L. and Borondo, F. and Scialchi, G. and Roncaglia, A. J. and Carlo, G. G. and Wisniacki, D. A.},
     month = {August},
     title = {Quantum reservoir complexity by the {Krylov} evolution approach},
   journal = {Phys. Rev. A},
    volume = {110},
    number = {2},
      year = {2024},
     pages = {022446},
       url = {https://link.aps.org/doi/10.1103/PhysRevA.110.022446},
       doi = {10.1103/physreva.110.022446}, 
}

@ARTICLE{CIN24,
     author = {{\v C}indrak, S. and Donvil, B. and L{\"{u}}dge, K. and Jaurigue, L. C.},
      month = {Jan},
      title = {Enhancing the performance of quantum reservoir computing and solving the time-complexity problem by artificial memory restriction},
    journal = {Phys. Rev. Research},
     volume = {6},
       year = {2024},
      pages = {013051},
  publisher = {American Physical Society}, 
        url = {https://link.aps.org/doi/10.1103/PhysRevResearch.6.013051},
        doi = {10.1103/physrevresearch.6.013051},
issue={1},
numpages={11},
}

@ARTICLE{CIN25,
     author = {{\v C}indrak, S. and L{\"{u}}dge, K. and Jaurigue, L. C.},
      month = {Nov},
      title = {From Krylov complexity to observability: Capturing phase space dimension with applications in quantum reservoir computing},
    journal = {Phys. Rev. Res.},
     volume = {7},
       year = {2025},
      pages = {L042039},
  publisher = {American Physical Society},
       issn = {2643-1564},
        url = {https://link.aps.org/doi/10.1103/m9kq-wrln},
        doi = {10.1103/m9kq-wrln}
}

@ARTICLE{CIN25a,
     author = {{\v C}indrak, S. and Jaurigue, L. C. and L{\"{u}}dge, K.},
      month = {Nov},
      title = {Engineering quantum reservoirs through Krylov complexity, expressivity, and observability},
    journal = {Phys. Rev. Res.},
     volume = {7},
       year = {2025},
      pages = {043190},
  publisher = {American Physical Society},
       issn = {2643-1564},
        url = {https://link.aps.org/doi/10.1103/84f3-63mz},
        doi = {10.1103/84f3-63mz}
}

@ARTICLE{MUJ23,
    author = {Mujal, P. and Mart{\'{\i}}nez-Pe{\~{n}}a, R. and Giorgi, G. L. and Soriano, M. C. and Zambrini, R.},
     title = {Time-series quantum reservoir computing with weak and projective measurements},
   journal = {npj Quantum Inf.},
    volume = {9},
    number = {1},
      year = {2023},
     pages = {16},
      issn = {2056-6387},
       url = {https://doi.org/10.1038/s41534-023-00682-z}, 
}

@ARTICLE{FRA24,
     author = {Franceschetto, G. and P{\l}odzie{\'{n}}, M. and Lewenstein, M. and Ac{\'{\i}}n, A. and Mujal, P.},
      month = {November},
      title = {Harnessing quantum back-action for time-series processing},
    journal = {arxiv:2411.03979}, 
       year = {2024}, 
        url = {http://arxiv.org/abs/2411.03979}, 
}

@ARTICLE{CHE24b,
    author = {Chepuri, R. and Amzalag, D. and Antonsen, T. M. and Girvan, M.},
     month = {06},
     title = {Hybridizing traditional and next-generation reservoir computing to accurately and efficiently forecast dynamical systems},
   journal = {Chaos},
    volume = {34},
    number = {6},
      year = {2024},
     pages = {063114},
      issn = {1054-1500},
       url = {https://doi.org/10.1063/5.0206232},
       doi = {10.1063/5.0206232}, 
}

@ARTICLE{PFE22,
    author = {Pfeffer, P. and Heyder, F. and Schumacher, J.},
     month = {September},
     title = {Hybrid quantum-classical reservoir computing of thermal convection flow},
   journal = {Phys. Rev. Research},
    volume = {4},
      year = {2022},
     pages = {033176},
       doi = {10.1103/physrevresearch.4.033176},
issue={3},
numpages={14},
}

@ARTICLE{AHM24,
    author = {Ahmed, O. and Tennie, F. and Magri, L.},
     month = {November},
     title = {Prediction of chaotic dynamics and extreme events: {A} recurrence-free quantum reservoir computing approach},
   journal = {Phys. Rev. Research},
    volume = {6},
    number = {4},
      year = {2024},
     pages = {043082},
      issn = {2643-1564},
       url = {https://link.aps.org/doi/10.1103/PhysRevResearch.6.043082},
       doi = {10.1103/physrevresearch.6.043082}
}

@ARTICLE{KOB24,
    author = {Kobayashi, K. and Fujii, K. and Yamamoto, N.},
     month = {November},
     title = {Feedback-{Driven} {Quantum} {Reservoir} {Computing} for {Time}-{Series} {Analysis}},
   journal = {PRX Quantum},
    volume = {5},
    number = {4},
      year = {2024},
     pages = {040325},
       url = {https://link.aps.org/doi/10.1103/PRXQuantum.5.040325},
       doi = {10.1103/prxquantum.5.040325}
}

@ARTICLE{KOB24a,
     author = {Kobayashi, S. and Tran, Q. H. and Nakajima, K.},
      month = {Aug},
      title = {Extending echo state property for quantum reservoir computing},
    journal = {Phys. Rev. E},
     volume = {110},
       year = {2024},
      pages = {024207},
  publisher = {American Physical Society},
        url = {https://link.aps.org/doi/10.1103/PhysRevE.110.024207},
        doi = {10.1103/physreve.110.024207},
issue={2},
numpages={15},
}

@ARTICLE{KOB25,
    author = {Kobayashi, K. and Motome, Y.},
     month = jun,
     title = {Quantum reservoir probing: {An} inverse paradigm of quantum reservoir computing for exploring quantum many-body physics},
   journal = {SciPost Phys.},
    volume = {18},
    number = {6},
      year = {2025},
     pages = {198},
      issn = {2542-4653},
       url = {https://www.scipost.org/SciPostPhys.18.6.198},
       doi = {10.21468/scipostphys.18.6.198},
shorttitle={Quantum reservoir probing},
}

@ARTICLE{KOB26,
    author = {Kobayashi, K. and Motome, Y.},
     month = jan,
     title = {Edge of {Many}-{Body} {Quantum} {Chaos} in {Quantum} {Reservoir} {Computing}},
   journal = {Phys. Rev. Lett.},
    volume = {136},
    number = {4},
      year = {2026},
     pages = {040602},
       url = {https://link.aps.org/doi/10.1103/j2qj-vwcl},
       doi = {10.1103/j2qj-vwcl}
}

@ARTICLE{CHE20b,
    author = {Chen, J. and Nurdin, H. I. and Yamamoto, N.},
     month = {August},
     title = {Temporal Information Processing on Noisy Quantum Computers},
   journal = {Phys. Rev. Applied},
    volume = {14},
      year = {2020},
     pages = {024065},
       doi = {10.1103/physrevapplied.14.024065},
issue={2},
numpages={22},
}

@ARTICLE{SAN24,
    author = {Sannia, A. and Mart{\'{\i}}nez-Pe{\~{n}}a, R. and Soriano, M. C. and Giorgi, G. L. and Zambrini, R.},
     month = {March},
     title = {Dissipation as a resource for {Quantum} {Reservoir} {Computing}},
   journal = {Quantum},
    volume = {8},
      year = {2024},
     pages = {1291},
       url = {https://quantum-journal.org/papers/q-2024-03-20-1291/},
       doi = {10.22331/q-2024-03-20-1291}, 
}

@ARTICLE{DOM23,
    author = {Domingo, L. and Carlo, G. and Borondo, F.},
     month = {May},
     title = {Taking advantage of noise in quantum reservoir computing},
   journal = {Sci. Rep.},
    volume = {13},
    number = {1},
      year = {2023},
     pages = {8790},
      issn = {2045-2322},
       url = {https://www.nature.com/articles/s41598-023-35461-5},
       doi = {10.1038/s41598-023-35461-5}, 
}

@ARTICLE{FRY23,
    author = {Fry, D. and Deshmukh, A. and Chen, S. Y. and Rastunkov, V. and Markov, V.},
     month = {November},
     title = {Optimizing quantum noise-induced reservoir computing for nonlinear and chaotic time series prediction},
   journal = {Sci. Rep.},
    volume = {13},
    number = {1},
      year = {2023},
     pages = {19326},
      issn = {2045-2322},
       url = {https://www.nature.com/articles/s41598-023-45015-4},
       doi = {10.1038/s41598-023-45015-4}, 
}

@ARTICLE{MAR21,
    author = {Mart{\'{\i}}nez-Pe{\~{n}}a, R. and Giorgi, G. L. and Nokkala, J. and Soriano, M. C. and Zambrini, R.},
     title = {Dynamical {Phase} {Transitions} in {Quantum} {Reservoir} {Computing}},
   journal = {Phys. Rev. Lett.},
    volume = {127},
    number = {10},
      year = {2021},
     pages = {100502},
      issn = {0031-9007, 1079-7114},
       url = {https://link.aps.org/doi/10.1103/PhysRevLett.127.100502},
       doi = {10.1103/physrevlett.127.100502}
}

@ARTICLE{DAM12,
    author = {Dambre, J. and Verstraeten, D. and Schrauwen, B. and Massar, S.},
     title = {Information processing capacity of dynamical systems},
   journal = {Sci. Rep.},
    volume = {2},
      year = {2012},
     pages = {514},
       doi = {10.1038/srep00514}
}

@ARTICLE{BAU24,
     author = {Bauwens, I. and Harkhoe, K. and Gooskens, E. and Bienstman, P. and Verschaffelt, G. and Van der Sande, G.},
      month = {Jul},
      title = {Combining a passive spatial photonic reservoir computer with a semiconductor laser increases its nonlinear computational capacity},
    journal = {Opt. Express},
     volume = {32},
     number = {14},
       year = {2024},
      pages = {24328--24345},
  publisher = {Optica Publishing Group},
        url = {https://opg.optica.org/oe/abstract.cfm?URI=oe-32-14-24328},
        doi = {10.1364/oe.518654}, 
}

@ARTICLE{HUE22a,
    author = {H{\"{u}}lser, T. and K{\"{o}}ster, F. and L{\"{u}}dge, K. and Jaurigue, L. C.},
     title = {Deriving task specific performance from the information processing capacity of a reservoir computer},
   journal = {Nanophotonics},
    volume = {12},
    number = {5},
      year = {2023},
     pages = {937},
       doi = {10.1515/nanoph-2022-0415}
}

@ARTICLE{INU17,
    author = {Inubushi, M. and Yoshimura, K.},
     title = {Reservoir Computing Beyond Memory-Nonlinearity Trade-off},
   journal = {Sci. Rep.},
    volume = {7},
      year = {2017},
     pages = {10199}
}

@ARTICLE{RAT25,
     author = {Rathor, S. K. and Ziegler, M. and Schumacher, J.},
      month = {Jan},
      title = {Asymmetrically connected reservoir networks learn better},
    journal = {Phys. Rev. E},
     volume = {111},
       year = {2025},
      pages = {015307},
  publisher = {American Physical Society},
        url = {https://link.aps.org/doi/10.1103/PhysRevE.111.015307},
        doi = {10.1103/physreve.111.015307},
issue={1},
numpages={8},
}

@ARTICLE{HUA26,
    author = {Huang, J. and Wang, T. and L{\"{u}}dge, K. and Han, Y. and Xiang, S. and Hao, Y.},
     title = {Stochastic modeling of microcavity laser-based photonic reservoir computing: An information processing capacity perspective},
   journal = {Neural Netw.},
    volume = {197},
      year = {2026},
     pages = {108383},
      issn = {0893-6080},
       url = {https://www.sciencedirect.com/science/article/pii/S089360802501264X},
       doi = {https://doi.org/10.1016/j.neunet.2025.108383}, 
}

@ARTICLE{MAR20,
    author = {Mart{\'{\i}}nez-Pe{\~{n}}a, R. and Nokkala, J. and Giorgi, G. L. and Zambrini, R. and Soriano, M. C.},
     month = sep,
     title = {Information {Processing} {Capacity} of {Spin}-{Based} {Quantum} {Reservoir} {Computing} {Systems}},
   journal = {Cogn. Comput.},
    volume = {15},
    number = {5},
      year = {2020},
     pages = {1440--1451},
      issn = {1866-9964},
       url = {https://doi.org/10.1007/s12559-020-09772-y},
       doi = {10.1007/s12559-020-09772-y}, 
}

@ARTICLE{GOE25,
    author = {G{\"{o}}tting, N. and Wilksen, S. and Steinhoff, A. and Lohof, F. and Gies, C.},
     month = dec,
     title = {Connection between {Memory} {Performance} and {Optical} {Absorption} in {Quantum} {Reservoir} {Computing}},
   journal = {Phys. Rev. Lett.},
    volume = {135},
    number = {24},
      year = {2025},
       url = {https://link.aps.org/doi/10.1103/vp79-8t1l},
       doi = {10.1103/vp79-8t1l}
}

@ARTICLE{LOR63,
    author = {Lorenz, E. N.},
     title = {Deterministic nonperiodic flow},
   journal = {J. Atmos. Sci.},
    volume = {20},
      year = {1963},
     pages = {130},
       doi = {10.1175/1520-0469(1963)020%3C0130:DNF%3E2.0.CO;2},
nota={gf:Lorenz-System},
}

@ARTICLE{WER19,
    author = {Wernecke, H. and S{\'{a}}ndor, B. and Gros, C.},
     month = sep,
     title = {Chaos in time delay systems, an educational review},
   journal = {Phys. Rep.},
    series = {Chaos in time delay systems, an educational review},
    volume = {824},
      year = {2019},
     pages = {1--40},
      issn = {0370-1573},
       url = {https://www.sciencedirect.com/science/article/pii/S0370157319302601},
       doi = {10.1016/j.physrep.2019.08.001}
}

@ARTICLE{KUB21,
    author = {Kubota, T. and Takahashi, H. and Nakajima, K.},
     title = {Unifying framework for information processing in stochastically driven dynamical systems},
   journal = {Phys. Rev. Research},
    volume = {3},
    number = {4},
      year = {2021},
     pages = {043135},
       doi = {10.1103/physrevresearch.3.043135}
}

@ARTICLE{KOE21,
    author = {K{\"{o}}ster, F. and Yanchuk, S. and L{\"{u}}dge, K.},
     month = {April},
     title = {Insight into delay based reservoir computing via eigenvalue analysis},
   journal = {J. Phys. Photonics},
    volume = {3},
    number = {2},
      year = {2021},
     pages = {024011},
       doi = {10.1088/2515-7647/abf237}, 
}

@ARTICLE{KOE22,
    author = {K{\"{o}}ster, F. and Yanchuk, S. and L{\"{u}}dge, K.},
     title = {Master memory function for delay-based reservoir computers with single-variable dynamics},
   journal = {IEEE Trans. Neural Netw. Learn. Syst.},
    volume = {35},
    number = {6},
      year = {2024},
     pages = {7712},
       doi = {10.1109/tnnls.2022.3220532},
}

@ARTICLE{GOL20,
    author = {Goldmann, M. and K{\"{o}}ster, F. and L{\"{u}}dge, K. and Yanchuk, S.},
     title = {Deep Time-Delay Reservoir Computing: Dynamics and Memory Capacity},
   journal = {Chaos,},
    volume = {30},
    number = {9},
      year = {2020},
     pages = {093124},
       doi = {doi: 10.1063/5.0017974},
	   
}

@ARTICLE{KOE20a,
    author = {K{\"{o}}ster, F. and Ehlert, D. and L{\"{u}}dge, K.},
     title = {Limitations of the recall capabilities in delay based reservoir computing systems},
   journal = {Cogn. Comput.},
    volume = {15},
      year = {2020},
     pages = {1419},
      note = {Springer, ISSN 1866-9956},
      issn = {1866-9956},
       doi = {10.1007/s12559-020-09733-5}
}

@ARTICLE{BAU22c,
    author = {Bauwens, I. and Van der Sande, G. and Bienstman, P. and Verschaffelt, G.},
     title = {Using photonic reservoirs as preprocessors for deep neural networks},
   journal = {Front. Phys.},
    volume = {10},
      year = {2022},
      issn = {2296-424X},
       url = {https://www.frontiersin.org/articles/10.3389/fphy.2022.1051941},
       doi = {10.3389/fphy.2022.1051941}, 
}

@ARTICLE{MAC77,
    author = {Mackey, M. C. and Glass, L.},
     title = {Oscillation and Chaos in Physiological Control Systems},
   journal = {Science},
    volume = {197},
    number = {4300},
      year = {1977},
     pages = {287-289},
       doi = {10.1126/science.267326}, 
}

@ARTICLE{ZHU25,
    author = {Zhu, C. and Ehlers, P. J. and Nurdin, H. I. and Soh, D.},
     title = {Practical and Scalable Quantum Reservoir Computing},
   journal = {Phys. Rev. Research},
    volume = {7},
      year = {2025},
     pages = {023290},
       doi = {10.1103/wsyq-jyxd},
primaryclass={quant-ph},
}

@ARTICLE{JAE04,
    author = {Jaeger, H. and Haas, H.},
     title = {{Harnessing Nonlinearity: Predicting Chaotic Systems and Saving Energy in Wireless Communication}},
   journal = {Science},
    volume = {304},
    number = {5667},
      year = {2004},
     pages = {78--80},
       doi = {10.1126/science.1091277}
}

@article{brunWMTestonQCs2008,
  title = {Test of Weak Measurement on a Two- or Three-Qubit Computer},
  author = {Brun, Todd A. and Di{\'o}si, Lajos and Strunz, Walter T.},
  year = {2008},
  month = mar,
  journal = {Physical Review A},
  volume = {77},
  number = {3},
  pages = {032101},
  issn = {1050-2947, 1094-1622},
  doi = {10.1103/PhysRevA.77.032101},
  urldate = {2024-05-23},
  copyright = {http://link.aps.org/licenses/aps-default-license},
  langid = {english}
}

@misc{bowlesFindingBenchmarks2024,
      title={Better than classical? The subtle art of benchmarking quantum machine learning models}, 
      author={Joseph Bowles and Shahnawaz Ahmed and Maria Schuld},
      year={2024},
      eprint={2403.07059},
      archivePrefix={arXiv},
      primaryClass={quant-ph},
      url={https://arxiv.org/abs/2403.07059}, 
}

@misc{SCH21a,
	title = {Supervised quantum machine learning models are kernel methods},
	url = {http://arxiv.org/abs/2101.11020},
	doi = {10.48550/arXiv.2101.11020}, 
	publisher = {arXiv},
	author = {Schuld, Maria}, 
	year = {2021},
	note = {arXiv:2101.11020 [quant-ph]},
}

@article{HAV19,
	title = {Supervised learning with quantum-enhanced feature spaces},
	volume = {567},
	issn = {1476-4687},
	url = {https://www.nature.com/articles/s41586-019-0980-2},
	doi = {10.1038/s41586-019-0980-2},
	number = {7747},
	journal = {Nature},
	author = {Havlíček, V. and Córcoles, A. D. and Temme, K. and Harrow, A. W. and Kandala, A. and Chow, J. M. and Gambetta, J. M.},
	month = mar,
	year = {2019},
	note = {Publisher: Nature Publishing Group},
	pages = {209--212},
	
}

@article{chenQRConNISQ2020,
  title = {Temporal {{Information Processing}} on {{Noisy Quantum Computers}}},
  author = {Chen, Jiayin and Nurdin, Hendra I. and Yamamoto, Naoki},
  year = 2020,
  month = aug,
  journal = {Physical Review Applied},
  volume = {14},
  number = {2},
  pages = {024065},
  doi = {10.1103/PhysRevApplied.14.024065},
  urldate = {2024-07-26}, 
}

@ARTICLE{JAU24,
     author = {Jaurigue, L. C. and L{\"{u}}dge, K.},
      title = {Reducing reservoir computer hyperparameter dependence by external timescale tailoring},
    journal = {Neuromorph. Comput. Eng.},
     volume = {4},
     number = {1},
       year = {2024},
      pages = {014001},
  publisher = {IOP Publishing},
        url = {https://dx.doi.org/10.1088/2634-4386/ad1d32},
        doi = {10.1088/2634-4386/ad1d32}, 
}

@ARTICLE{TAN19c,
    author = {Tanaka, G. and Yamane, T. and H{\'{e}}roux, J. B. and Nakane, R. and Kanazawa, N. and Takeda, S. and Numata, H. and Nakano, D. and Hirose, A.},
     title = {Recent advances in physical reservoir computing: A review},
   journal = {Neural Netw.},
    volume = {115},
      year = {2019},
     pages = {100--123},
      issn = {0893-6080},
       doi = {https://doi.org/10.1016/j.neunet.2019.03.005}, 
}

@ARTICLE{APP11,
    author = {Appeltant, L. and Soriano, M. C. and Van der Sande, G. and Danckaert, J. and Massar, S. and Dambre, J. and Schrauwen, B. and Mirasso, C. R. and Fischer, I.},
     title = {Information processing using a single dynamical node as complex system},
   journal = {Nat. Commun.},
    volume = {2},
      year = {2011},
     pages = {468},
       doi = {10.1038/ncomms1476}, 
}

@ARTICLE{BRU13a,
    author = {Brunner, D. and Soriano, M. C. and Mirasso, C. R. and Fischer, I.},
     month = {January},
     title = {Parallel photonic information processing at gigabyte per second data rates using transient states},
   journal = {Nat. Commun.},
    volume = {4},
      year = {2013},
     pages = {1364},
       doi = {10.1038/ncomms2368}
}

@INBOOK{LUE26,
     author = {L{\"{u}}dge, K. and Jaurigue, L. C.},
     editor = {te Vrugt, M.},
      title = {Time-multiplexed Reservoir Computing with Semiconductor Laser Systems},
  booktitle = {Artificial Intelligence and Intelligent Matter: Nanoscience, Soft Matter, Philosophy},
       year = {2026},
      pages = {371--388},
  publisher = {Springer Nature Switzerland},
    address = {Cham},
       isbn = {978-3-032-04129-6},
        url = {https://doi.org/10.1007/978-3-032-04129-6_18},
        doi = {10.1007/978-3-032-04129-6_18}, 
}

@ARTICLE{FUM25,
    author = {Fumagalli, L. and L{\"{u}}dge, K. and de Wiljes, J. and Haario, H. and Jaurigue, L. C.},
     month = nov,
     title = {Data-Driven Performance Measures using Global Properties of Attractors for Testing Black-Box Surrogate Models of Chaotic Systems},
   journal = {Chaos: An Interdisciplinary Journal of Nonlinear Science},
    volume = {35},
    number = {11},
      year = {2025},
     pages = {113121},
      issn = {1054-1500},
       url = {https://doi.org/10.1063/5.0283424},
       doi = {10.1063/5.0283424}
}

@ARTICLE{BIS95,
  author={Bishop, C. M.},
  journal={Neural Comput.}, 
  title={Training with Noise is Equivalent to Tikhonov Regularization}, 
  year={1995},
  volume={7},
  number={1},
  pages={108-116},
  doi={10.1162/neco.1995.7.1.108}}

@ARTICLE{AUS25,
    author = {Austin, M. and Nakajima, K.},
     month = {09},
     title = {Quantization induced memory-nonlinearity transfer: Implications of analog-to-digital conversion in reservoir computing},
   journal = {Chaos},
    volume = {35},
    number = {9},
      year = {2025},
     pages = {093107},
      issn = {1054-1500},
       url = {https://doi.org/10.1063/5.0273403},
       doi = {10.1063/5.0273403},
}

@article{SWI25, 
  title = {Fading ergodicity meets maximal chaos},
  author = {\ifmmode \acute{S}\else \'{S}\fi{}wi\ifmmode \mbox{\c{e}}\else \c{e}\fi{}tek, Rafa\l{} and \L{}yd\ifmmode \dot{z}\else \.{z}\fi{}ba, Patrycja and Vidmar, Lev},
  journal = {Phys. Rev. B},
  volume = {111},
  issue = {18},
  pages = {184203},
  numpages = {14},
  year = {2025},
  month = {May},
  publisher = {American Physical Society},
  doi = {10.1103/PhysRevB.111.184203},
  url = {https://link.aps.org/doi/10.1103/PhysRevB.111.184203}
}
\end{document}